\documentclass[iop]{emulateapj}

\usepackage{natbib}
\usepackage{epsfig}
\usepackage{epstopdf}
\begin{document}

\title{A Uniform Search for Secondary Eclipses of Hot Jupiters in \emph{Kepler} Q2 Lightcurves}

\author{J. L. Coughlin\altaffilmark{1,4} and M. L\'opez-Morales\altaffilmark{2,3}}
\journalinfo{Accepted for Publication in the Astronomical Journal}

\altaffiltext{1}{Department of Astronomy, New Mexico State University, P.O. Box 30001, MSC 4500, Las Cruces, New Mexico 88003-8001}
\altaffiltext{2}{Institut de Ci\`encies de l'Espai (CSIC-IEEC), Campus UAB, Facultat de Ci\`encies, Torre C5, parell, 2a pl, E-08193 Bellaterra, Barcelona, Spain}
\altaffiltext{3}{Carnegie Institution of Washington, Department of Terrestrial Magnetism, 5241 Broad Branch Road NW, Washington, DC 20015-1305, USA; Visiting Investigator}
\altaffiltext{4}{NSF Graduate Research Fellow}
\email{jlcough@nmsu.edu}

\begin{abstract}
We present in this paper the results of searching the $Kepler$ Q2 public dataset for the secondary eclipses of 76 hot Jupiter planet candidates from the list of 1,235 candidates published by \citet{Borucki2011}. This search has been performed by modeling both the $Kepler$ pre-search data conditioned light curves and new light curves produced via our own photometric pipeline. We derive new stellar and planetary parameters for each system, while calculating robust errors for both. We find 16 systems with 1-2$\sigma$, 14 systems with 2-3$\sigma$, and 6 systems with $>$3$\sigma$ confidence level secondary eclipse detections in at least one light curve produced via the Kepler pre-search data conditioned light curve or our own pipeline, however, results can vary depending on the light curve modeled and whether eccentricity is allowed to vary or not. We estimate false alarm probabilities of 31\%, 10\%, and 6\% for the 1-2$\sigma$, 2-3$\sigma$, and $>$3$\sigma$ confidence intervals respectively. Comparing each secondary eclipse result to theoretical expectations, we find that the majority of detected planet candidates emit more light than expected due to thermal blackbody emission in the optical $Kepler$ bandpass, and present a trend of increasing excess emission with decreasing maximum effective planetary temperature. These results agree with previously published optical secondary eclipse data for other hot Jupiters. We explore modeling biases, significant planetary albedos, non-LTE or other thermal emission, significant internal energy generation, and mis-identification of brown dwarfs, low-mass stars, or stellar blends as possible causes of both the excess emission and its correlation with expected planetary temperature. Although we find no single cause is able to explain all of the planet candidates, significant planetary albedos, with a general trend of increasing planetary albedos with decreasing atmospheric temperatures, is able to explain most of the systems. Identifying systems that we deem likely to be low-mass stars or stellar blends, we estimate an 11\% false positive rate in the current $Kepler$ planet candidate sample of hot Jupiters. We also establish robust upper limits on the eclipse depth for the remaining systems, and find that the emission of a significant fraction of these systems is consistent with the planets having very low albedos, i.e., at least 30\% of all systems have $A_{g}$~$<$~0.3 at 1$\sigma$ confidence levels. This result augments the current number of constrained exoplanetary albedos and extends the sample of low albedo determinations to planets with temperatures as low as 1200~K. Finally, we note that continued observations with the $Kepler$ spacecraft, and improved techniques for the removal of systematic noise in the $Kepler$ data, are needed to better characterize these systems.
\end{abstract}


\section{Introduction}

\label{introsec}

Measuring the secondary eclipses of transiting exoplanets at optical wavelengths is a powerful tool for probing their atmospheres, in particular their albedos, brightness temperatures, and energy redistribution factors. The $Kepler$ mission has recently uncovered over a thousand new transiting planet candidates \citep{Borucki2011}, which provide an unprecedented and uniform sample of high photometric precision light curves among which secondary eclipse signals can be detected.

In the past decade, many surprising discoveries regarding the atmospheric properties of hot Jupiters have been made. For example, many hot Jupiters appear to have temperature inversions, with numerous proposed explanations, but no definitive evidence for exactly which physical processes are involved \citep{Hubeny2003,Fortney2006,Burrows2007,Fortney2008,Spiegel2009,Zahnle2009,Knutson2010,Madhusudhan2010}. Other results have found that the atmospheric composition of different planets vary significantly, or that they present a wide range of heat circulation efficiencies between their day and night sides \citep[see][and references therein]{Baraffe2010}.

Most of the observations yielding to those discoveries have been done in the mid-infrared (3.6 - 24 $\mu$m) with $Spitzer$. Observations at shorter wavelengths are more scarce, especially in the visible, but most of them point towards the predominance of very low geometric albedo, ($A_g$ $<$ 0.3 at the 3$\sigma$ level upper limits), atmospheres in hot Jupiters \citep{Charbonneau1999,Leigh2003a,Leigh2003b,Rodler2008,Rodler2010,Rowe2008,Alonso2009a,Alonso2009b,Alonso2010,Snellen2009,Christiansen2010,Welsh2010,KippingBakos2011a,KippingSpiegel2011,Desert2011a,Desert2011b,Langford2011}, in contrast to the up to $A_g=0.5$ albedos observed in the colder gas giants in our Solar System \citep{Karkoschka1994}. Those results are in fair agreement with early theoretical models \citep[e.g.,][]{Marley1999,Sudarsky2000,Seager2000}, which predict significant absorption of the incident stellar radiation in the visible by sodium and potassium, followed by re-emission in the infrared. Other molecules, such as TiO, VO, and HS, have also been suggested as possible strong optical absorbers \citep[e.g.,][]{Hubeny2003,Fortney2008,Zahnle2009}.

However, three recent studies suggest higher geometric albedos for two planets. \citet{Berdyugina2011} have published a value of $A_g$ = 0.28$\pm$0.16 for HD 189733b via polarized reflected light\footnote{We note that \citep{Wiktorowicz2009} reported a non-detection of polarized light from this planet, and placed an upper limit to the polarimetric modulation of the exoplanet at $\Delta$P $<$ 7.9$\times$10$^{-5}$.}, while \citet{KippingBakos2011a} and \citet{Demory2011} respectively suggest albedos of $A_g$ = 0.38$\pm$0.12 and $A_g$ = 0.32$\pm$0.03 for Kepler-7b based on measurements of the emission of the planet with $Kepler$ during secondary eclipse. Some plausible explanations for such high albedos include Rayleigh scattering and the presence of clouds or hazes in the atmospheres of those planets \citep{Demory2011}. Also, in the case of the hottest planets, some amount of thermal emission could be contributing to the measured emission levels in the reddest edge of the observed visible wavelength windows (e.g., at $\lambda \sim$ 0.8 $\mu$m).

Although new theoretical work is necessary to determine the cause of apparently high albedos in some hot Jupiters, the key answer to whether these results are typical or not relies on more observations, since the current discussions are based on a statistically insufficient sample of only three planets. The purpose of this work is to significantly increase that sample by searching for the emission of hot Jupiters among the publicly available $Kepler$ light curves of planet candidates reported by \citet{Borucki2011}. Given the photometric precision of the $Kepler$ data and the wavelength coverage of the $Kepler$ passband (0.4 - 0.9 $\mu$m), these datasets provide unprecedented quality data to detect the secondary eclipses of those planets in the visible and statistically determine the albedos of hot Jupiters. Furthermore, as \citet{Borucki2011} do no explicitly state how they modeled their light curves or obtained their parameters, a re-modeling of the data will perform an independent test on the methods they employed.

In addition to providing estimations of the planetary albedo, measuring the timing and duration of the secondary eclipse, when coupled with the primary eclipse, can directly measure the orbital eccentricity of a system \citep[e.g.,][]{Knutson2007}. Also, if there is a significant flux contrast between the day and night side of the planet, one may be able to measure the varying amount of emitted light by the planet in the light curve, and directly measure the day-to-night contrast ratio \citep{Harrington2006,Knutson2007}. Even a robust upper limit on the eclipse depth can narrow the range of possible planetary albedos and yield useful information on the statistics of exoplanetary albedos.

In \S\ref{datasec} we present our target selection criteria and describe how we reprocess the $Kepler$ light curves from the pixel-level data. In \S\ref{modelsec} we describe how we model the data using the JKTEBOP code, and obtain robust errors on all parameters while accounting for potential systematic noise. We present our derived physical parameters of both the planet candidates and their host stars in \S\ref{paramsec}, and in \S\ref{trendssec} we examine possible trends in our results. In \S\ref{indivsec} we discuss individual candidates of interest, and finally in \S\ref{concsec} we summarize our findings and examine possible future directions for the study of this sample.

\section{Observational Data}
\label{datasec}

The first step of our analysis consisted of selecting a set of planet candidates suitable for secondary eclipse detection among the 1,235 planet candidates published by \citet{Borucki2011}. We made a pre-selection of potentially detectable objects using the planetary and stellar parameters listed in Table 2 of \citet{Borucki2011}, choosing only those systems with $P$ $<$ 5 days and $R_p$ $>$ 0.5 $R_{J}$, after estimating that planets with longer orbital periods and smaller radii are too cool, too small, or too far away from their host star to produce deep enough secondary eclipse signals to be detectable by $Kepler$, even in the most extreme albedo conditions (i.e., $A_g$ = 1.0). Our secondary eclipse depth estimations also account for the amount of stellar irradiation, given the effective temperature of the stars reported by \citet{Borucki2011}. The result is a list of 76 candidates.

The next step consisted of an inspection of the $Kepler$ light curves of those 76 targets. The analysis of \citet{Borucki2011} uses the first four months of $Kepler$
observations, which include quarters Q0, Q1, and Q2. However, significantly discrepant systematic noise patterns exist between the light curves from different quarters, which result in additional noise when all the data are combined. Therefore, we decided to use only the data from Q2, which alone contains continuous 90-day observation coverage and is well suited for our search.

We modeled two different light curves for each target: the Presearch Data Conditioned (PDC) light curve, and our own generated light curves that we produced using the pixel-level data and our own photometric pipeline. In the remainder of the paper we refer to this second analysis as the CLM pipeline. As detailed in \citet{Jenkins2010a}, the first step in creating the PDC light curves was correcting the pixel level data for bias pattern noise, dark current, gain, non-linearity, cosmic rays, shutter smearing, pixel-to-pixel sensitivity, and other pixel-level effects. The calibrated pixels were then run through a Photometric Analysis (PA) that measures and subtracts background flux, and sums up pixels within a photometric aperture for each star, creating the PA light curve. The size of those apertures are defined such that it is supposed to maximize the mean signal-to-noise for each star. The PA light curves were then subjected to Pre-Search Data Conditioning which attempts to remove systematic effects due to temperature, focus, pointing, and other effects by correlating with ancillary engineering data. The PDC module also corrects for any sudden jumps in the data, for example due to sudden pointing changes or pixel sensitivities due to cosmic ray hits, as well as removes excess flux in the photometric aperture due to crowding.

During our inspection of the PDC data we noted that, despite the thorough analysis detailed by \citet{Jenkins2010a}, many of the PDC light curves produced by the $Kepler$ PDC pipeline still contain significant systematic trends at a level of couple percent variation, an effect that can significantly hinder the detection of secondary eclipses and phase brightness variations. Upon thorough examination of the pixel-level data, PA, and PDC light curves, we concluded that the majority of the trends correlate with and are due to the 0.1-0.5 pixel centroid position drift experienced by the majority of the target stars each quarter. This drift is principally due to Differential Velocity Aberration (DVA), where the amount of stellar aberration introduced by the spacecraft's velocity varies over the large field of view, resulting in the shifting of stellar positions on the detector as large as 0.6 pixels over a 90 day period \citep{Jenkins2010b}. Spacecraft pointing error only accounts for 0.05 pixels of the total movement \citep{Jenkins2010b}. This drift in the stellar position causes light from the wings of each star's point spread function to enter and leave the optimal photometric aperture at different rates, resulting in a flux variance of several percent over each quarter. To remove those effects we re-analyzed the $Kepler$ pixel-level data using the CLM pipeline.

The CLM photometric pipeline starts with the calibrated pixel-level data, with background flux removed from each pixel. As the majority of target stars are well-isolated, we simply summed up the flux for every pixel that was downloaded from the spacecraft for each star. We find that this removes the majority of long-term systematic variations due to DVA, usually producing light curves with significantly less systematic noise than the similarly produced PA light curves. Usually the only time summing up all the pixels produced more systematic noise is when there was significant crowding in the field by comparably bright stars, but we find this only affects a small fraction of the selected transiting planet candidates, and note that crowding can still significantly affect PA photometry as well. 

Even after minimizing the amount of light variation within the aperture, pixel-to-pixel sensitivity, both in spectral response as well as quantum efficiency, and intrapixel variations, still produced significant systematics. We cut out areas of significant systematic variation, which principally occur around BJD 2455015 and BJD 2455065, due to a safe mode event and a pointing tweak. Then, we performed a correlation-based Principal Component Analysis (PCA) \citep{Murtagh1987} on the pixel level data, and subtract out the first three PCA components, thus removing the majority of major systematic noise, which is still principally correlated with the large position drift. We have checked and verified that this does not remove or significantly modify the transit signal for any system. We then fit a B\'ezier curve \citep{Kahaner1989} to the data, performed a 3-sigma rejection, re-fit a B\'ezier curve, and then divided by the fit. This iterative procedure does an excellent job of removing any possible remaining low-frequency systematic features in the data without removing or affecting the transits or real high-frequency stellar variations. We then, for only a few systems, removed one or two points that were extremely significant outliers. For one system, KOI 433.01, we removed a single transit that was clearly from another long-period companion in the system. For both the $Kepler$ PDC and CLM light curves, we finally subtracted a linear trend to ensure the light curves were completely flat before modeling (see \S\ref{modelsec}).

We would like to note that, as a possible technique of removing systematic noise, we also attempted to directly solve for a pixel mask that would account for pixel-to-pixel variation in quantum efficiency and spectral response. Every image in the time series was multiplied by this pixel mask, whose values ranged from 0 to 1, and then the pixel fluxes summed to produce a corrected light curve. We used an asexual genetic algorithm, similar to that presented in \citet{Coughlin2011}, to solve for the values of each pixel in the pixel mask that produced corrected light curves with a minimum amount of systematic noise, defined via various methods. We found that the technique was very successful at removing nearly all systematic noise from the light curves. However, depending on the minimization criteria selected, we found that the algorithm was prone to over-correct the light curve, and remove features due to real astrophysical phenomena. As well, even when it did appear to remove the systematic trends and not the real astrophysical signatures, it was difficult to tell, unlike with the PCA analysis, whether or not the solution had a real physical basis. Thus, we decided not to employ this technique in our analysis. However, with more work or a better understanding of the systematic noise sources, it might become a viable means for removing systematic noise from $Kepler$, and possibly other, light curves.

\begin{figure*}
\centering
\epsfig{width=0.7\linewidth,file=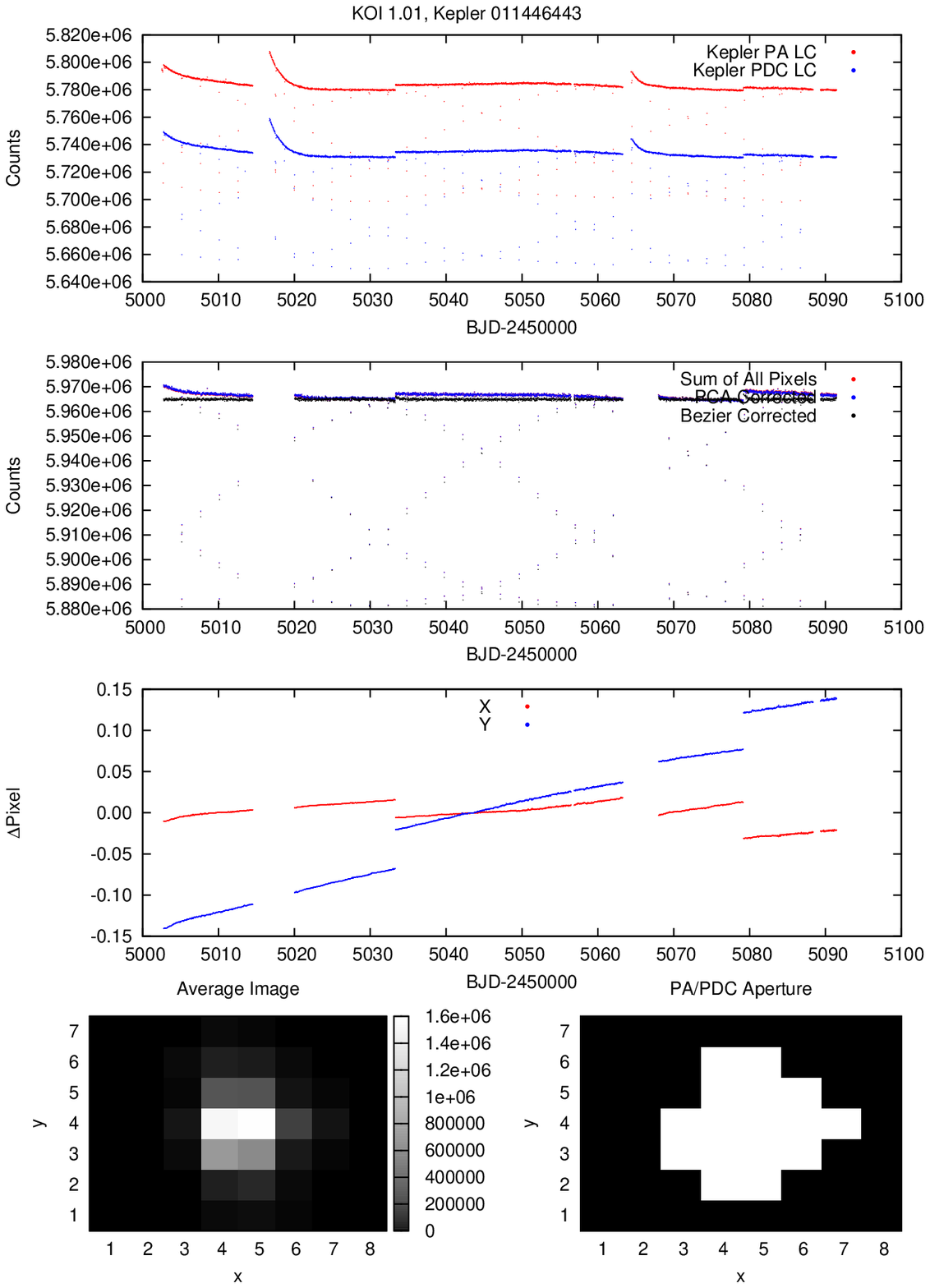} 
\caption{Plots of the light curves, centroid positions, pixel-level images, and photometric apertures used in the $Kepler$ PA and PDC reduction, for all initial 76 candidate systems. The top panel for each system shows the Q2 $Kepler$ PA and PDC light curves. The next panel shows our CLM pipeline reduced light curves, including those from simply summing all the pixels in each frame, applying PCA correction, and then applying a Bezier correction, with the most severe systematic noise regions cut out. The next panel shows the flux-weighted relative centroid movement in both X and Y over Q2, using all pixels in the frame, again with the most severe systematic noise regions cut out. The bottom left panel is the average image of all the frames over the quarter. The bottom right panel shows the photometric aperture used in the $Kepler$ PA and PDC light curve reduction, where only white pixels were counted and summed. Only the first plot, Figure~\ref{candlcs}.1, is shown in the text for guidance. Figures~\ref{candlcs}.1-\ref{candlcs}.76 are available in the online version of the Journal.}
\label{candlcs}
\end{figure*}

In Figure~\ref{candlcs} we plot the $Kepler$ PA and PDC light curves, our CLM light curves, (including those from simply summing up all the pixels in each frame, applying the PCA correction, and then applying the Bezier correction), the centroid positions, averaged pixel-level image, and the aperture used in the $Kepler$ PA and PDC photometry, for each of the 76 candidate systems. Of the original 76 candidates, 36 were deemed unmodelable based on their PDC light curves, due to either strong systematics or intrinsic stellar variability with amplitudes on the order of, or greater than, the depth of the transits. The 36 discarded systems were KOI 1.01, 17.01, 20.01, 127.01, 128.01, 135.01, 183.01, 194.01, 203.01, 208.01, 214.01, 217.01, 254.01, 256.01, 552.01, 554.01, 609.01, 667.01, 767.01, 823.01, 882.01, 883.01, 895.01, 981.01, 1152.01, 1176.01, 1177.01, 1227.01, 1285.01, 1382.01, 1448.01, 1452.01, 1540.01, 1541.01, 1543.01, and 1546.01. In the case of our newly generated CLM pipeline light curves, 26 candidates turned out to be unmodelable, most of them due to stellar variability, as in the case of the PDC light curves above, or due to blends in the images resulting in significant light contamination of the target light curves. The 26 systems discarded in this case were KOI 102.01, 135.01, 194.01, 199.01, 208.01, 256.01, 552.01, 554.01, 609.01, 823.01, 882.01, 883.01, 895.01, 931.01, 961.02, 961.03, 981.01, 1152.01, 1177.01, 1227.01, 1285.01, 1382.01, 1448.01, 1452.01, 1540.01, and 1546.01. Thus, in total, there are 21 targets that have no modelable light curve from either analysis, (nearly all due to intrinsic stellar variability), 35 systems that have modelable light curves from both the $Kepler$ PDC data and our CLM reduction, and 55 systems that have at least one modelable light curve from either the $Kepler$ PDC data or our CLM analysis. In Table~\ref{tab1} we present the Kepler Object of Interest (KOI) number, $Kepler$ ID number, and host stars' $Kepler$ magnitude, effective temperature, surface gravity, and metallicity from the Kepler Input Catalog of each of the 55 modelable candidates.

\begin{deluxetable*}{rccccrcccc}
  \tablewidth{0pt}
  \tabletypesize{\scriptsize}
  \tablecaption{Modeled Systems and Their Host Star Properties}
  \tablecolumns{10}
  \tablehead{KOI & \emph{Kepler} ID & m$_{\rm kep}$ & $T_{\star}$ & $log~g$ & [Fe/H] & $M_{\star,\rm KIC}$ & $R_{\star, \rm KIC}$ & $M_{\star,\rm ISO}$ & $R_{\star, \rm ISO}$\\ & & & (K) & (cgs) & & ($M_{\sun}$) & ($R_{\sun}$) & ($M_{\sun}$) & ($R_{\sun}$)}
  \startdata
  1.01 & 011446443 & 11.338 & 5713 & 4.14 & -0.139 & 1.133$^{_{+1.726}}_{^{-0.681}}$ & 1.496$^{_{+1.400}}_{^{-0.727}}$ & 1.019$^{_{+0.060}}_{^{-0.053}}$ & 0.943$^{_{+0.074}}_{^{-0.060}}$ \\
2.01 & 010666592 & 10.463 & 6577 & 4.32 & 0.000 & 1.351$^{_{+2.059}}_{^{-0.812}}$ & 1.336$^{_{+1.251}}_{^{-0.649}}$ & 1.316$^{_{+0.078}}_{^{-0.075}}$ & 1.363$^{_{+0.115}}_{^{-0.113}}$ \\
5.01 & 008554498 & 11.665 & 5766 & 4.04 & 0.116 & 1.176$^{_{+1.720}}_{^{-0.706}}$ & 1.748$^{_{+1.574}}_{^{-0.846}}$ & 1.035$^{_{+0.060}}_{^{-0.055}}$ & 0.962$^{_{+0.077}}_{^{-0.065}}$ \\
10.01 & 006922244 & 13.563 & 6164 & 4.44 & -0.128 & 1.110$^{_{+1.718}}_{^{-0.666}}$ & 1.056$^{_{+1.002}}_{^{-0.510}}$ & 1.162$^{_{+0.074}}_{^{-0.066}}$ & 1.133$^{_{+0.109}}_{^{-0.093}}$ \\
13.01 & 009941662 & 9.958 & 8848 & 3.93 & -0.141 & 1.813$^{_{+2.807}}_{^{-1.088}}$ & 2.454$^{_{+2.329}}_{^{-1.186}}$ & 1.989$^{_{+0.055}}_{^{-0.053}}$ & 1.616$^{_{+0.021}}_{^{-0.022}}$ \\
17.01 & 010874614 & 13.000 & 5724 & 4.47 & 0.000 & 0.908$^{_{+1.353}}_{^{-0.538}}$ & 0.911$^{_{+0.842}}_{^{-0.443}}$ & 1.022$^{_{+0.060}}_{^{-0.054}}$ & 0.947$^{_{+0.075}}_{^{-0.061}}$ \\
18.01 & 008191672 & 13.369 & 5816 & 4.46 & 0.000 & 0.922$^{_{+1.447}}_{^{-0.551}}$ & 0.942$^{_{+0.862}}_{^{-0.454}}$ & 1.050$^{_{+0.062}}_{^{-0.059}}$ & 0.981$^{_{+0.081}}_{^{-0.071}}$ \\
20.01 & 011804465 & 13.438 & 6012 & 4.47 & -0.161 & 1.069$^{_{+1.637}}_{^{-0.639}}$ & 0.999$^{_{+0.933}}_{^{-0.483}}$ & 1.110$^{_{+0.071}}_{^{-0.061}}$ & 1.059$^{_{+0.100}}_{^{-0.080}}$ \\
64.01 & 007051180 & 13.143 & 5128 & 3.94 & -0.341 & 1.180$^{_{+1.737}}_{^{-0.694}}$ & 1.956$^{_{+1.729}}_{^{-0.936}}$ & 0.868$^{_{+0.048}}_{^{-0.047}}$ & 0.786$^{_{+0.046}}_{^{-0.042}}$ \\
97.01 & 005780885 & 12.885 & 5944 & 4.27 & 0.052 & 1.086$^{_{+1.705}}_{^{-0.649}}$ & 1.279$^{_{+1.170}}_{^{-0.616}}$ & 1.089$^{_{+0.067}}_{^{-0.060}}$ & 1.031$^{_{+0.094}}_{^{-0.077}}$ \\
102.01 & 008456679 & 12.566 & 5919 & 3.90 & -0.358 & 1.216$^{_{+1.862}}_{^{-0.727}}$ & 2.058$^{_{+1.922}}_{^{-0.994}}$ & 1.081$^{_{+0.067}}_{^{-0.061}}$ & 1.020$^{_{+0.092}}_{^{-0.075}}$ \\
127.01 & 008359498 & 13.938 & 5570 & 4.53 & 0.174 & 1.043$^{_{+1.581}}_{^{-0.634}}$ & 0.921$^{_{+0.850}}_{^{-0.436}}$ & 0.980$^{_{+0.056}}_{^{-0.053}}$ & 0.897$^{_{+0.066}}_{^{-0.055}}$ \\
128.01 & 011359879 & 13.758 & 5718 & 4.18 & 0.362 & 1.139$^{_{+1.700}}_{^{-0.688}}$ & 1.443$^{_{+1.284}}_{^{-0.699}}$ & 1.020$^{_{+0.061}}_{^{-0.054}}$ & 0.945$^{_{+0.076}}_{^{-0.061}}$ \\
144.01 & 004180280 & 13.698 & 4724 & 4.00 & 0.241 & 1.098$^{_{+1.680}}_{^{-0.662}}$ & 1.744$^{_{+1.568}}_{^{-0.838}}$ & 0.767$^{_{+0.053}}_{^{-0.055}}$ & 0.694$^{_{+0.049}}_{^{-0.052}}$ \\
183.01 & 009651668 & 14.290 & 5722 & 4.71 & -0.141 & 1.012$^{_{+1.536}}_{^{-0.608}}$ & 0.734$^{_{+0.678}}_{^{-0.353}}$ & 1.022$^{_{+0.062}}_{^{-0.054}}$ & 0.947$^{_{+0.077}}_{^{-0.062}}$ \\
186.01 & 012019440 & 14.952 & 5826 & 4.56 & 0.021 & 1.059$^{_{+1.660}}_{^{-0.632}}$ & 0.877$^{_{+0.819}}_{^{-0.414}}$ & 1.052$^{_{+0.063}}_{^{-0.058}}$ & 0.984$^{_{+0.083}}_{^{-0.070}}$ \\
188.01 & 005357901 & 14.741 & 5087 & 4.73 & 0.255 & 0.893$^{_{+1.353}}_{^{-0.542}}$ & 0.671$^{_{+0.619}}_{^{-0.318}}$ & 0.859$^{_{+0.046}}_{^{-0.048}}$ & 0.778$^{_{+0.043}}_{^{-0.042}}$ \\
195.01 & 011502867 & 14.835 & 5604 & 4.50 & -0.188 & 1.058$^{_{+1.580}}_{^{-0.640}}$ & 0.968$^{_{+0.862}}_{^{-0.469}}$ & 0.988$^{_{+0.059}}_{^{-0.051}}$ & 0.907$^{_{+0.071}}_{^{-0.053}}$ \\
196.01 & 009410930 & 14.465 & 5585 & 4.51 & 0.096 & 1.029$^{_{+1.574}}_{^{-0.620}}$ & 0.937$^{_{+0.842}}_{^{-0.450}}$ & 0.984$^{_{+0.056}}_{^{-0.053}}$ & 0.901$^{_{+0.067}}_{^{-0.054}}$ \\
199.01 & 010019708 & 14.879 & 6214 & 4.60 & 0.104 & 1.081$^{_{+1.641}}_{^{-0.650}}$ & 0.863$^{_{+0.797}}_{^{-0.415}}$ & 1.182$^{_{+0.076}}_{^{-0.070}}$ & 1.161$^{_{+0.113}}_{^{-0.100}}$ \\
201.01 & 006849046 & 14.014 & 5491 & 4.45 & 0.187 & 1.039$^{_{+1.628}}_{^{-0.620}}$ & 0.985$^{_{+0.920}}_{^{-0.465}}$ & 0.959$^{_{+0.053}}_{^{-0.052}}$ & 0.875$^{_{+0.059}}_{^{-0.052}}$ \\
202.01 & 007877496 & 14.309 & 5912 & 4.44 & 0.120 & 1.093$^{_{+1.667}}_{^{-0.654}}$ & 1.052$^{_{+0.952}}_{^{-0.516}}$ & 1.080$^{_{+0.064}}_{^{-0.060}}$ & 1.019$^{_{+0.088}}_{^{-0.075}}$ \\
203.01 & 010619192 & 14.141 & 5634 & 4.49 & 0.041 & 1.056$^{_{+1.592}}_{^{-0.636}}$ & 0.980$^{_{+0.915}}_{^{-0.473}}$ & 0.996$^{_{+0.060}}_{^{-0.051}}$ & 0.915$^{_{+0.072}}_{^{-0.054}}$ \\
204.01 & 009305831 & 14.678 & 5287 & 4.48 & -0.104 & 0.966$^{_{+1.447}}_{^{-0.577}}$ & 0.950$^{_{+0.854}}_{^{-0.460}}$ & 0.905$^{_{+0.054}}_{^{-0.045}}$ & 0.821$^{_{+0.054}}_{^{-0.042}}$ \\
214.01 & 011046458 & 14.256 & 5322 & 4.44 & 0.018 & 1.022$^{_{+1.510}}_{^{-0.612}}$ & 1.015$^{_{+0.890}}_{^{-0.488}}$ & 0.916$^{_{+0.052}}_{^{-0.048}}$ & 0.832$^{_{+0.054}}_{^{-0.046}}$ \\
217.01 & 009595827 & 15.127 & 5504 & 4.72 & 0.220 & 0.975$^{_{+1.448}}_{^{-0.584}}$ & 0.706$^{_{+0.649}}_{^{-0.330}}$ & 0.963$^{_{+0.054}}_{^{-0.053}}$ & 0.879$^{_{+0.060}}_{^{-0.053}}$ \\
229.01 & 003847907 & 14.720 & 5608 & 4.37 & 0.219 & 1.069$^{_{+1.616}}_{^{-0.645}}$ & 1.124$^{_{+1.041}}_{^{-0.533}}$ & 0.990$^{_{+0.056}}_{^{-0.052}}$ & 0.909$^{_{+0.067}}_{^{-0.054}}$ \\
254.01 & 005794240 & 15.979 & 3948 & 4.54 & 0.234 & 0.530$^{_{+0.810}}_{^{-0.322}}$ & 0.650$^{_{+0.602}}_{^{-0.307}}$ & 0.347$^{_{+0.168}}_{^{-0.118}}$ & 0.319$^{_{+0.124}}_{^{-0.090}}$ \\
356.01 & 011624249 & 13.807 & 5124 & 4.07 & -0.503 & 1.111$^{_{+1.623}}_{^{-0.665}}$ & 1.593$^{_{+1.466}}_{^{-0.750}}$ & 0.869$^{_{+0.048}}_{^{-0.049}}$ & 0.787$^{_{+0.045}}_{^{-0.043}}$ \\
412.01 & 005683743 & 14.288 & 5584 & 4.28 & -0.011 & 1.093$^{_{+1.667}}_{^{-0.654}}$ & 1.275$^{_{+1.153}}_{^{-0.625}}$ & 0.983$^{_{+0.058}}_{^{-0.053}}$ & 0.901$^{_{+0.068}}_{^{-0.054}}$ \\
421.01 & 009115800 & 14.995 & 5181 & 4.32 & -0.075 & 1.016$^{_{+1.500}}_{^{-0.610}}$ & 1.155$^{_{+1.053}}_{^{-0.547}}$ & 0.880$^{_{+0.051}}_{^{-0.047}}$ & 0.797$^{_{+0.049}}_{^{-0.042}}$ \\
433.01 & 010937029 & 14.924 & 5237 & 4.37 & 0.375 & 0.986$^{_{+1.476}}_{^{-0.589}}$ & 1.080$^{_{+0.971}}_{^{-0.523}}$ & 0.894$^{_{+0.052}}_{^{-0.046}}$ & 0.810$^{_{+0.052}}_{^{-0.041}}$ \\
611.01 & 006309763 & 14.022 & 6122 & 4.55 & -0.132 & 1.085$^{_{+1.599}}_{^{-0.645}}$ & 0.914$^{_{+0.852}}_{^{-0.429}}$ & 1.149$^{_{+0.069}}_{^{-0.066}}$ & 1.115$^{_{+0.102}}_{^{-0.092}}$ \\
667.01 & 006752502 & 13.826 & 4135 & 4.57 & 0.000 & 0.607$^{_{+0.917}}_{^{-0.364}}$ & 0.681$^{_{+0.597}}_{^{-0.326}}$ & 0.501$^{_{+0.126}}_{^{-0.165}}$ & 0.431$^{_{+0.114}}_{^{-0.121}}$ \\
684.01 & 007730747 & 13.831 & 5331 & 3.96 & 0.113 & 1.174$^{_{+1.743}}_{^{-0.704}}$ & 1.870$^{_{+1.718}}_{^{-0.873}}$ & 0.917$^{_{+0.053}}_{^{-0.048}}$ & 0.833$^{_{+0.054}}_{^{-0.045}}$ \\
760.01 & 011138155 & 15.263 & 5887 & 4.62 & 0.010 & 1.060$^{_{+1.611}}_{^{-0.634}}$ & 0.840$^{_{+0.801}}_{^{-0.403}}$ & 1.072$^{_{+0.063}}_{^{-0.061}}$ & 1.008$^{_{+0.087}}_{^{-0.076}}$ \\
767.01 & 011414511 & 15.052 & 5431 & 4.44 & 0.023 & 1.026$^{_{+1.615}}_{^{-0.625}}$ & 1.007$^{_{+0.943}}_{^{-0.485}}$ & 0.946$^{_{+0.050}}_{^{-0.052}}$ & 0.862$^{_{+0.053}}_{^{-0.052}}$ \\
801.01 & 003351888 & 15.001 & 5472 & 4.39 & 0.182 & 1.041$^{_{+1.590}}_{^{-0.632}}$ & 1.080$^{_{+1.000}}_{^{-0.510}}$ & 0.955$^{_{+0.054}}_{^{-0.054}}$ & 0.871$^{_{+0.059}}_{^{-0.054}}$ \\
809.01 & 003935914 & 15.530 & 5690 & 4.48 & -0.385 & 1.047$^{_{+1.518}}_{^{-0.630}}$ & 0.967$^{_{+0.877}}_{^{-0.454}}$ & 1.012$^{_{+0.061}}_{^{-0.053}}$ & 0.934$^{_{+0.075}}_{^{-0.059}}$ \\
813.01 & 004275191 & 15.725 & 5357 & 4.73 & -0.285 & 0.955$^{_{+1.428}}_{^{-0.576}}$ & 0.695$^{_{+0.641}}_{^{-0.328}}$ & 0.924$^{_{+0.052}}_{^{-0.050}}$ & 0.839$^{_{+0.054}}_{^{-0.047}}$ \\
830.01 & 005358624 & 15.224 & 4915 & 4.90 & 0.155 & 0.797$^{_{+1.176}}_{^{-0.478}}$ & 0.528$^{_{+0.481}}_{^{-0.250}}$ & 0.819$^{_{+0.047}}_{^{-0.052}}$ & 0.742$^{_{+0.041}}_{^{-0.049}}$ \\
838.01 & 005534814 & 15.311 & 5794 & 4.48 & -0.095 & 1.049$^{_{+1.596}}_{^{-0.619}}$ & 0.987$^{_{+0.913}}_{^{-0.471}}$ & 1.043$^{_{+0.059}}_{^{-0.057}}$ & 0.972$^{_{+0.078}}_{^{-0.068}}$ \\
840.01 & 005651104 & 15.028 & 4916 & 4.39 & -0.091 & 0.936$^{_{+1.379}}_{^{-0.556}}$ & 1.023$^{_{+0.953}}_{^{-0.481}}$ & 0.818$^{_{+0.048}}_{^{-0.052}}$ & 0.742$^{_{+0.042}}_{^{-0.049}}$ \\
843.01 & 005881688 & 15.270 & 5784 & 4.40 & 0.203 & 1.093$^{_{+1.650}}_{^{-0.655}}$ & 1.109$^{_{+0.972}}_{^{-0.531}}$ & 1.041$^{_{+0.060}}_{^{-0.057}}$ & 0.969$^{_{+0.079}}_{^{-0.068}}$ \\
897.01 & 007849854 & 15.257 & 5734 & 4.46 & 0.270 & 1.066$^{_{+1.597}}_{^{-0.637}}$ & 1.024$^{_{+0.934}}_{^{-0.497}}$ & 1.025$^{_{+0.060}}_{^{-0.056}}$ & 0.951$^{_{+0.075}}_{^{-0.064}}$ \\
908.01 & 008255887 & 15.113 & 5391 & 4.25 & 0.128 & 1.050$^{_{+1.591}}_{^{-0.636}}$ & 1.279$^{_{+1.197}}_{^{-0.618}}$ & 0.933$^{_{+0.052}}_{^{-0.051}}$ & 0.849$^{_{+0.053}}_{^{-0.050}}$ \\
913.01 & 008544996 & 15.198 & 5463 & 4.75 & -0.281 & 0.967$^{_{+1.521}}_{^{-0.589}}$ & 0.688$^{_{+0.644}}_{^{-0.332}}$ & 0.953$^{_{+0.052}}_{^{-0.053}}$ & 0.869$^{_{+0.057}}_{^{-0.053}}$ \\
931.01 & 009166862 & 15.272 & 5714 & 4.78 & 0.319 & 1.016$^{_{+1.547}}_{^{-0.617}}$ & 0.685$^{_{+0.624}}_{^{-0.330}}$ & 1.020$^{_{+0.061}}_{^{-0.053}}$ & 0.944$^{_{+0.075}}_{^{-0.061}}$ \\
961.02 & 008561063 & 15.920 & 4188 & 4.56 & 0.000 & 0.612$^{_{+0.888}}_{^{-0.368}}$ & 0.671$^{_{+0.609}}_{^{-0.315}}$ & 0.536$^{_{+0.117}}_{^{-0.157}}$ & 0.461$^{_{+0.114}}_{^{-0.123}}$ \\
961.03 & 008561063 & 15.920 & 4188 & 4.56 & 0.000 & 0.626$^{_{+0.955}}_{^{-0.377}}$ & 0.679$^{_{+0.621}}_{^{-0.325}}$ & 0.539$^{_{+0.115}}_{^{-0.155}}$ & 0.463$^{_{+0.113}}_{^{-0.123}}$ \\
1176.01 & 003749365 & 15.715 & 4601 & 4.69 & 0.376 & 0.746$^{_{+1.115}}_{^{-0.453}}$ & 0.648$^{_{+0.614}}_{^{-0.309}}$ & 0.736$^{_{+0.053}}_{^{-0.078}}$ & 0.663$^{_{+0.053}}_{^{-0.083}}$ \\
1419.01 & 011125936 & 15.507 & 5848 & 4.46 & -0.292 & 1.075$^{_{+1.614}}_{^{-0.645}}$ & 0.999$^{_{+0.920}}_{^{-0.474}}$ & 1.059$^{_{+0.065}}_{^{-0.059}}$ & 0.992$^{_{+0.087}}_{^{-0.072}}$ \\
1459.01 & 009761199 & 15.692 & 4060 & 4.40 & 0.098 & 0.646$^{_{+1.001}}_{^{-0.389}}$ & 0.830$^{_{+0.781}}_{^{-0.393}}$ & 0.444$^{_{+0.138}}_{^{-0.159}}$ & 0.384$^{_{+0.115}}_{^{-0.113}}$ \\
1541.01 & 004840513 & 15.189 & 6164 & 4.53 & 0.068 & 1.100$^{_{+1.685}}_{^{-0.657}}$ & 0.954$^{_{+0.850}}_{^{-0.465}}$ & 1.163$^{_{+0.074}}_{^{-0.068}}$ & 1.134$^{_{+0.108}}_{^{-0.095}}$ \\
1543.01 & 005270698 & 14.985 & 5821 & 4.54 & -0.240 & 1.044$^{_{+1.587}}_{^{-0.618}}$ & 0.912$^{_{+0.870}}_{^{-0.433}}$ & 1.052$^{_{+0.061}}_{^{-0.059}}$ & 0.983$^{_{+0.080}}_{^{-0.071}}$ \\

  \enddata
  \label{tab1}
\end{deluxetable*}

We now highlight a few systems to illustrate the different types of systematic and stellar noise in the $Kepler$ data, and differences between the PDC and CLM light curves. For KOI 17.01, (see Figure~\ref{candlcs}.6), the PA and PDC light curves both show a $\sim$1.7\% systematic variation over the quarter, as the PA/PDC aperture does not encompass many pixels in the wing of the PSF that contain significant signal, and the star experiences a $\sim$0.4 pixel drift over the quarter. In contrast, the raw pixel-summed CLM light curve shows only a $\sim$0.54\% systematic variation, and the PCA-corrected and Bezier-corrected CLM light curves show virtually no remaining systematic noise. For KOI 102.01 and 199.01, (see Figures~\ref{candlcs}.11 and \ref{candlcs}.22), there is a close companion star that causes the CLM photometry to produce much worse light curves than the PDC data. In the case of KOI 102.01, significant systematics are introduced in the CLM light curve from the movement of the companion in and out of the frame, given the $\sim$0.3 pixel drift over the quarter, and as well the extra third light causes the transits of the primary to be damped out. In the case of KOI 199.01, the companion is an eclipsing binary, and its light curve is imposed on top of that of the transiting system. Significant systematics are still present in the PA data in these two cases, but much less so than the CLM data, and they appear to be removed in the PDC data. In the cases of KOI 256.01 and KOI 1452.01, (see Figures~\ref{candlcs}.32 and \ref{candlcs}.71), the stars exhibit clear high-frequency variations at the same level as the transits, possibly due to stellar pulsation or rapid rotation and star spots. Note that the CLM pipeline does not remove the stellar signal because it is high-frequency and intrinsic to the system, and also that for KOI 256.01 additional long-term systematic noise is present in the $Kepler$ PA and PDC data due to the small aperture they employ and the $\sim$0.35 pixel drift, but does not exist in the CLM data.

\section{Light Curve Modeling}
\label{modelsec}

We used the JKTEBOP eclipsing binary modeling code \citep{Southworth2004a,Southworth2004b}, which is based on the EBOP code \citep{Etzel1981,Popper1981}, to model both the $Kepler$ PDC light curves and our own CLM pipeline light curves for the 55 modelable systems. In short, JKTEBOP\footnote{For more information on JKTEBOP, see http://www.astro.keele.ac.uk/jkt/codes/jktebop.html} models the projection of each star as a biaxial ellipsoid and calculates light curves by numerical integration of concentric circles over each star, and is well-suited to modeling detached eclipsing binaries or transiting extrasolar planets. We modeled each light curve first fixing the eccentricity to $e=0$, and then leaving it as a free parameter. The reason for leaving $e$ as a free parameter is that, even though systems with $P$ $<$ 5 days are generally expected to be circularized, additional bodies in the system or other evolutionary effects can perturb their orbits. Indeed, at the time of this writing, $\sim$36\% of currently known transiting planets with $P$ $<$ 5 days have a measured non-zero eccentricity \citep{exoplanetwebsite}. Therefore, since we are performing a blind search for secondary eclipses, restricting the search to only circular orbits might result in detection biases. The results between fixing $e=0$ and letting it vary can sometimes vary significantly, as shown at the end of this section.

For both cases of either fixing $e=0$ or letting it vary, we also simultaneously solved for the orbital period of the system, $P$, time of primary transit minimum, $T_{0}$, the inclination of the orbit, $i$, e$\cdot$cos($\omega$) and e$\cdot$sin($\omega$), where $\omega$ is the longitude of periastron, the planet-to-star surface brightness ratio, $J$, the sum of the fractional radii, $r_{sum}$, the planet-to-star radii ratio, $k$, and the out-of-eclipse (baseline) flux. (We note for clarity that the relation between $J$, $k$, and the planet-to-star luminosity ratio, $L_{r}$, is $L_{r}$ = $k^{2}J$.) To account for any potential brightness variations with phase, we also multiply the planet's luminosity, $L_{p}$, by a factor of one plus a sinusoidal curve, so that

\begin{equation}
  L_{p}(T) = L_{p} + A_{L_{p}} \cdot sin\left(\frac{2\pi(T-T_{0})}{P} - \frac{\pi}{2}\right)
\end{equation}

\noindent where $L_{p}(T)$ is the planet's luminosity at a given observed time, $T$, and $A_{L_{p}}$, for which we solve, is the amplitude of the sinusoidal curve. Note that we have fixed the period of this sine wave to the orbital period of each system, and fixed the reference zero phase so that the maximum amplitude peak coincides with the center of the secondary eclipse. Although there has been at least one case of a measured planetary brightness phase curve having its maximum offset from secondary eclipse in the infrared \citep{Knutson2007}, many optical observations indicate planetary brightness phase curve maximums coincident with the secondary eclipse \citep{Borucki2009,Snellen2009,Welsh2010,Bonomo2011}. We note that a value of $A_{L_{p}}$ = 0.0 implies no brightness variations with phase. A value of $A_{L_{p}}$ = 0.2 implies the planet is 20\% brighter at phase 0.5, when the day-side is visible, and 20\% fainter at phase 1.0, when the night-side is visible, compared to phases 0.25 and 0.75. A value of $A_{L_{p}}$ = 1.0 implies a perfectly dark night-side. Negative values of $A_{L_{p}}$ would imply a brighter night-side than day-side, which is not physically expected, but allowed for in the code so as not to introduce any bias towards positive values of $A_{L_{p}}$. Note also that $J$ is allowed to be both positive and negative so as not to introduce any bias towards positive values of $J$, and thus false detections. 

In both cases we assumed a quadratic limb-darkening law for the stars, and fixed coefficients to the values found by \citet{Sing2010} for the $Kepler$ bandpass, using the estimated stellar effective temperatures, surface gravities, and metallicities. We also set the values of the gravity darkening coefficients to those derived by \citet{Claret2000}, based on the effective temperature of the stars. Even though we fix the limb and gravity darkening coefficients, they are computed from stellar models and have an associated uncertainty when compared to reality. As the choice of these coefficients can affect the determination of other system parameters, their uncertainty must be taken into account in the error analysis. \citet{Claret2008} determined this uncertainty to be $\sim$10\%, and thus we allowed the values of the limb and gravity darkening coefficients to vary over a range of $\pm$10\% during the error estimation analysis, described below.

Finally, as pointed out by \citet{Kipping2010}, sparse sampling times, such as the 29.4244 minute sampling of the $Kepler$ long-cadence data, can significantly alter the morphological shape of a transit light curve and result in erroneous planetary parameter estimations if the effect is not taken into account. Thus, we instructed JKTEBOP to integrate the models over 29.4244 minutes, composed of 10 separate sub-intervals centered on each observed data point, to account for this effect. 

We derived error estimates using three error analysis techniques implemented in JKTEBOP: Monte Carlo, Bootstrapping, and Residual Permutation, but chose to adopt the parameter errors estimated by this last technique as it has been shown to best account for the effect of systematic noise in transit light curves \citep{Jenkins2002}. While Monte Carlo and Bootstrapping tend to underestimate errors in the presence of systematic noise, those two techniques have traditionally been chosen over Residual Permutation because in the latter one can only refit the data as many times as available data points. This poses a problem for most ground based transit light curves, which typically have only a couple hundred points, but for $Kepler$ Q2 data, which contains almost 5000 data points over a 90-day interval for the long-cadence data, and nearly 30 times more for the short-cadence data, the method is not statistically limited and therefore best suited to derive robust errors.

\begin{figure*}
\centering
\epsfig{width=0.75\linewidth,file=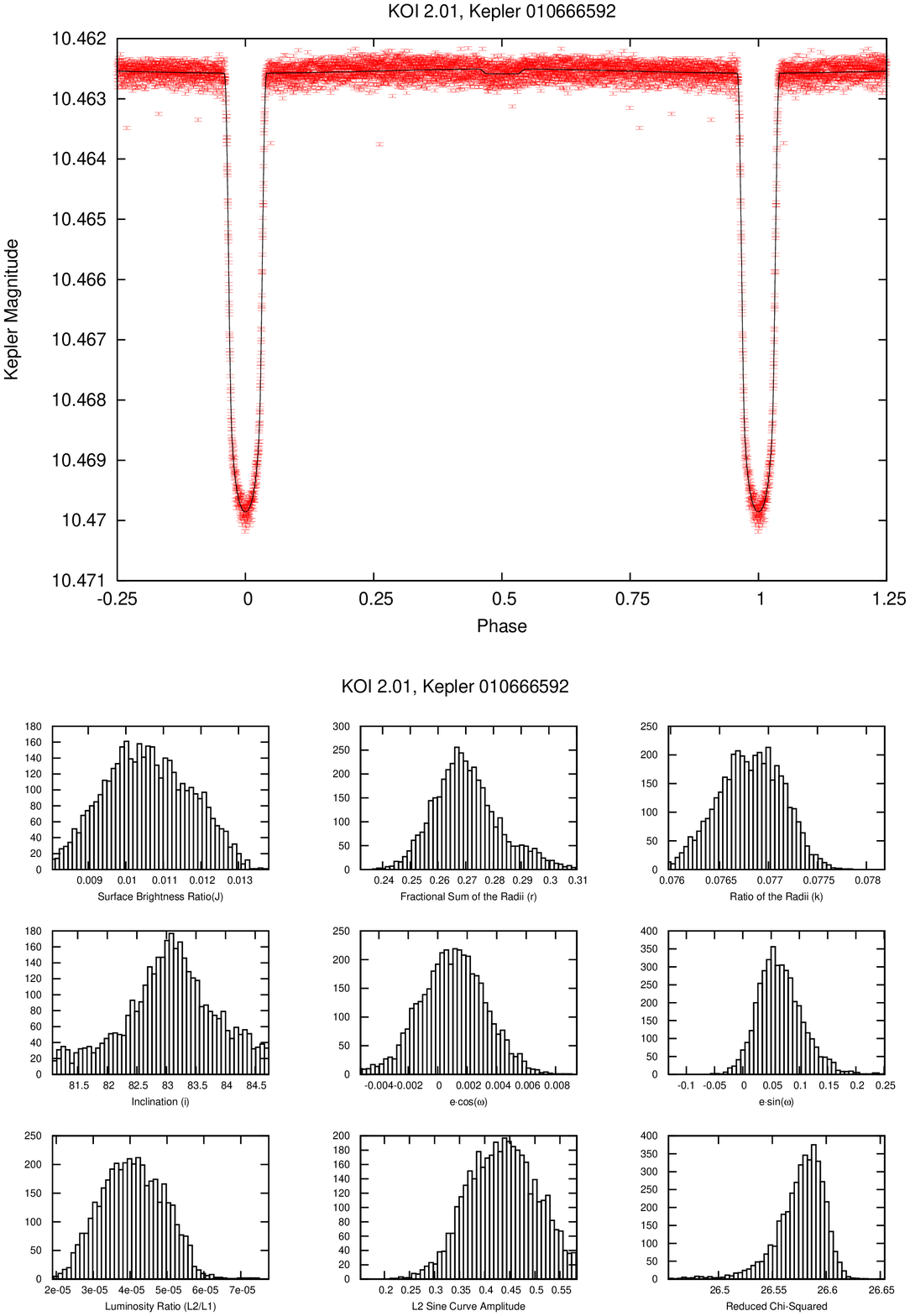}
\caption{Plots of the phased light curves of the 40 systems produced from the $Kepler$ PDC photometric pipeline, shown with our best model fits, allowing eccentricity to vary, and histograms of the resulting parameter distributions from the error analysis. Only the first plot, Figure~\ref{keplcs}.1, is shown in the text for guidance. Figures~\ref{keplcs}.1-\ref{keplcs}.40 are available in the online version of the Journal.}
\label{keplcs}
\end{figure*}

\begin{figure*}
\centering
\epsfig{width=0.75\linewidth,file=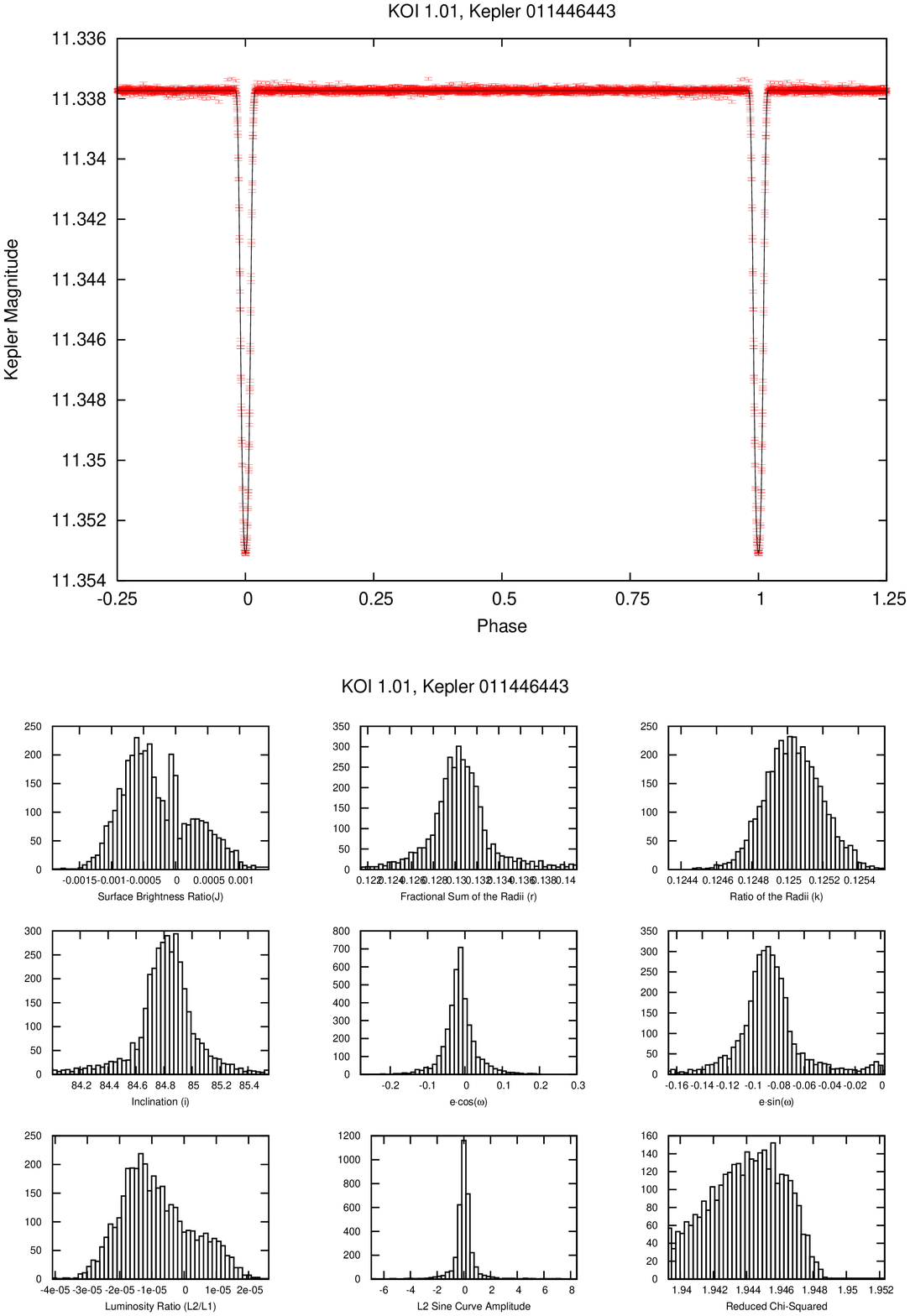}
\caption{Plots of the phased light curves of the 50 systems produced using our CLM photometric pipeline, shown with our best model fits, allowing eccentricity to vary, and histograms of the resulting parameter distributions from the error analysis. Only the first plot, Figure~\ref{ourlcs}.1, is shown in the text for guidance. Figures~\ref{ourlcs}.1-\ref{ourlcs}.50 are available in the online version of the Journal.}
\label{ourlcs}
\end{figure*}

In Figure~\ref{keplcs} we plot the resulting phased light curves, with the corresponding best-fit model light curve when allowing eccentricity to vary, along with histograms of the parameter distributions from the error analysis, for the 40 modelable systems with $Kepler$ PDC light curves. In Figure~\ref{ourlcs} we do the same, but for the 50 modelable systems with CLM light curves. In Table~\ref{modelresultstab} we list the median values for all the modeling parameters, for both sets of light curves, and for both fixing $e=0$ and allowing it to vary, along with their determined asymmetric, Gaussian, 1$\sigma$ errors.

As a result of our light curve modeling we find, when fixing $e=0$, 9 secondary eclipse detections at the 1-2$\sigma$ level, 3 detections at the 2-3$\sigma$ level and 4 detections at the $>$3$\sigma$ level in the PDC light curves. In the CLM light curves, we find 11 secondary eclipse detections at the 1-2$\sigma$ level, 4 detections at the 2-3$\sigma$ level and 4 detections at the $>$3$\sigma$ level. In the case of allowing eccentricity to vary we find 18 detections at 1-2$\sigma$ level, 3 detections at 2-3$\sigma$ level, and 4 detections at $>$3$\sigma$ level in the PDC light curves. In the CLM light curves we find 10 detections at 1-2$\sigma$ level, 10 detections at 2-3$\sigma$ level, and 5 detections at $>$3$\sigma$ level. Each set of results has been used independently in the statistical study of candidate emission parameters described in Sections~\ref{paramsec} and \ref{trendssec}. Examining both sets of light curves, and both $e=0$ and $e$ allowed to vary, we find 16 systems with 1-2$\sigma$, 14 systems with 2-3$\sigma$, and 6 systems with $>$3$\sigma$ confidence level secondary eclipse detections in at least one light curve. It is more difficult to quantify the number of systems that have certain level detections among multiple light curves and eccentricity constraints, given that not all systems had both PDC and CLM light curves and that eccentric systems may not be detected in the non-eccentric model, and is best left to the discussion of individual systems in \S\ref{indivsec}. Additionally, examining the 35 systems that had both modelable PDC and CLM light curves, we find that for the PDC light curves the average reduced $\chi^{2}$ value is 4.86, while for the CLM light curves it is 2.48, and that on average, each system's CLM light curve has a 27\% lower reduced $\chi^{2}$ value compared to the PDC light curve.

We also note that there were significant detections of negative $J$ values for some systems. When fixing eccentricity to zero, for the PDC light curves, we find 3 detections of negative $J$ at the 1-2$\sigma$ level, but none at higher significance. In the CLM light curves, we find 4 detections of negative $J$ at the 1-2$\sigma$ level, but none at higher significance. In the case of allowing eccentricity to vary, for the PDC light curves, we find 4 detections of negative $J$ at the 1-2$\sigma$ level, and 1 detection at the 2-3$\sigma$ level, but none at higher significance. In the CLM light curves we find 4 detections of negative $J$ at the 1-2$\sigma$ level, 1 detection at the 2-3$\sigma$ level, and 1 detection at the $>$3$\sigma$ level. Since there is no known physical mechanism to increase the flux of the system when the planet passes behind the host star, these detections are obviously spurious. Since there is no bias towards or preference for positive or negative $J$ values in the modeling code, and assuming that the Kepler data does not suffer from systematics that preferentially result in either decrements or increments in flux that span expected secondary eclipse durations, statistically speaking we must have as many false detections of positive $J$ values, or secondary eclipses, for as many detections of negative $J$ values we have, per each confidence interval. Thus, when fixing eccentricity to zero, for PDC light curves, we estimate our false alarm probabilities as 33\% for 1-2$\sigma$ detections, and 0\% for $>$2$\sigma$ detections. For CLM light curves, we estimate a 36\% false alarm probability for 1-2$\sigma$ detections, and 0\% for $>$2$\sigma$ detections. When allowing eccentricity to vary, for PDC light curves we estimate false alarm probabilities of 22\% for 1-2$\sigma$ detections, 33\% for 2-3$\sigma$ detections, and 0\% for $>$3$\sigma$ detections. For the CLM light curves, we estimate a 40\% false alarm probability for 1-2$\sigma$ detections, 10\% for 2-3$\sigma$ detections, and 20\% for $>$3$\sigma$ detections. Although we are dealing with small number statistics and the uncertainties on the determined false alarm probabilities are large, we note that these roughly agree with what we would statistically expect for each confidence interval quoted, i.e., a 1$\sigma$ detection has a formal 31.73\% false alarm probability by definition, though allowing eccentricity to vary does appear to induce false detections at $\sim$1.5 times greater frequency. Combining all the results from each light curve type and eccentricity parameter, we can generalize our false alarm probabilities to 31\%, 10\%, and 6\% for the 1-2$\sigma$, 2-3$\sigma$, and $>$3$\sigma$ confidence intervals respectively.

\section{Derivation of Stellar and Planetary Parameters}
\label{paramsec}

The secondary eclipse detections presented in the previous section allow, for the first time, to make a statistically significant analysis of the emission properties of exoplanet candidates at visible wavelengths, specifically in the $\sim$0.4 - 0.9 $\mu m$ $Kepler$ passband.

In this section we first revise the parameters of the host stars necessary to derive the physical properties of the planets and then compute physical and atmospheric parameters for each planet candidate, i.e., the brightness, equilibrium, and maximum effective temperatures, radii, and albedos, using both the originally reported and revised stellar parameter values. A detailed statistical study of the properties for the planets based on those parameters is presented in \S\ref{trendssec}.

\subsection{Stellar Parameters}

The Kepler Input Catalog (KIC) provides estimates of the effective temperature, surface gravity, and radius for all the host stars in our sample. Those parameters have been derived from a combination of broad and narrow-band photometry \citep{Brown2011}, although it has been also recognized that some of the parameter values in the KIC might contain significant errors.  As explained in detail by \citet{Brown2011}, the majority of approximately Sun-like stars in the KIC have effective temperatures that only disagree by 200 K or less from the temperature values of a control sample derived by other methods. However, for stars significantly more massive or less massive than the Sun, i.e., with $T_{\star} \gtrsim$ 9000 K and $T_{\star} \lesssim$ 4000 K, where $T_{\star}$ is the effective temperature of the star, the temperatures in the KIC can suffer from large systematics and are not reliable. As well, in the case of the derived stellar radii, $R_{\star}$, the values reported in the KIC are derived from statistical relations between the values obtained for $T_{\star}$, the surface gravity of the stars, $log~g$, and the luminosity, $L_{\star}$ (see sections 7 and 8 of \citet{Brown2011}, for more details). Therefore, if any of those parameters are systematically off, (such as $log~g$, which has an associated error of $\pm$0.4 dex), the values derived for $R_{\star}$ will be erroneous.

All the host stars in our sample have KIC effective temperature estimates between 4000 and 9000 K, so we have assumed that those values are accurate within the errors. From those temperatures we recomputed the radius and mass of the stars via interpolation of up-to-date stellar evolutionary models by \citet{Bertelli2008} for $M_{\star} \leq 1.4 M_{\sun}$ and \citet{Siess2000} for $M_{\star} >$ 1.4 $M_{\sun}$. In the models we have assumed that all the stars are nearly coeval, with an age of $\sim$1 Gyr and therefore on the main sequence, have abundances similar to the Sun, i.e., Z = 0.017, Y = 0.26, and have a mixing length of $\alpha$ = 1.68. 

The errors in those parameters have been estimated by recomputing the mass and radius of each star 10,000 times, each time adding random Gaussian noise to the underlying variables and examining the 1$\sigma$ spread of the resulting distribution. In the error estimations using the KIC values, we assumed an error of $\pm$0.4 dex for $log~g$, as reported by \citet{Brown2011}. In the error estimations using the model isochrones we assume an error of $\pm$200 K for $T_{\star}$, as reported by \citet{Brown2011}. The values for the mass and radius of each star as computed from the KIC, labeled ``KIC'', and via interpolation of the stellar isochrones, labeled ``ISO'', along with their estimated errors, are listed in Table~\ref{tab1}.

\subsection{Planetary Parameters}

\begin{figure*}
\centering
\begin{tabular}{cc}
  \epsfig{width=0.5\linewidth,file=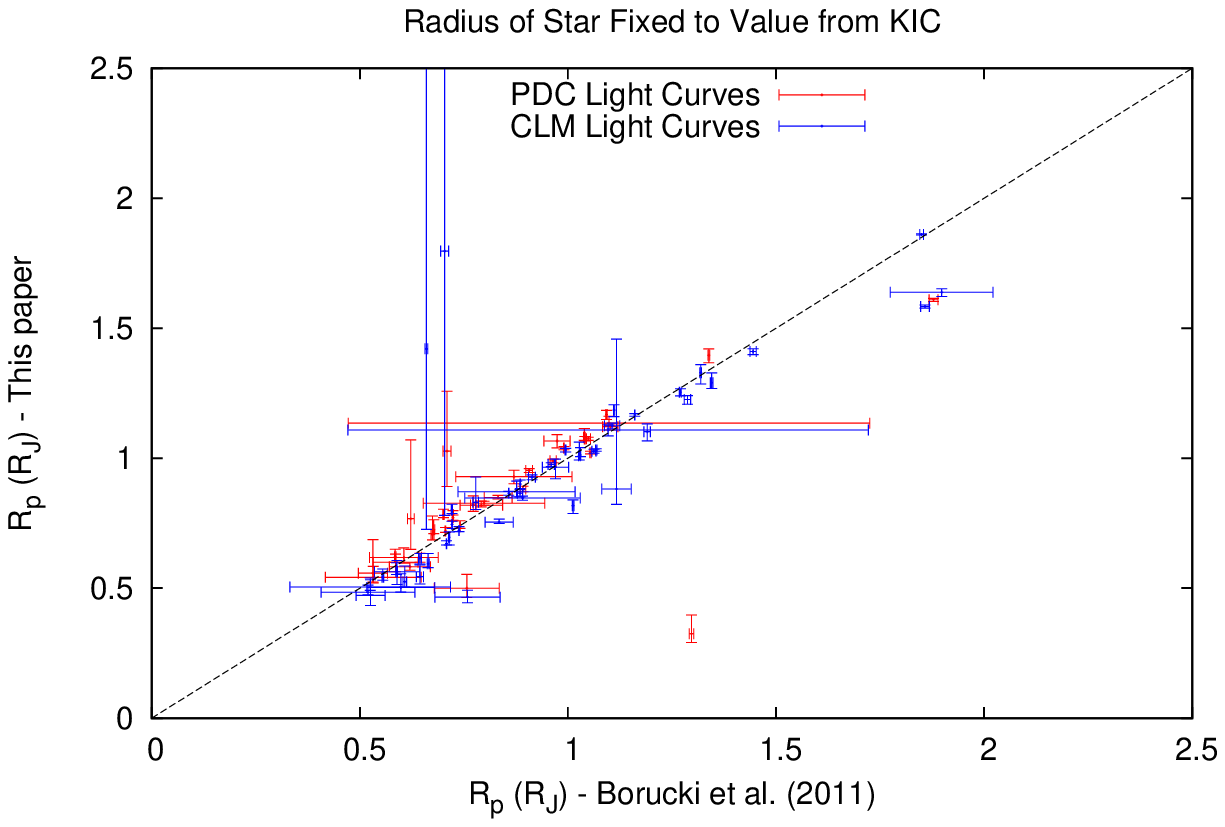}&
  \epsfig{width=0.5\linewidth,file=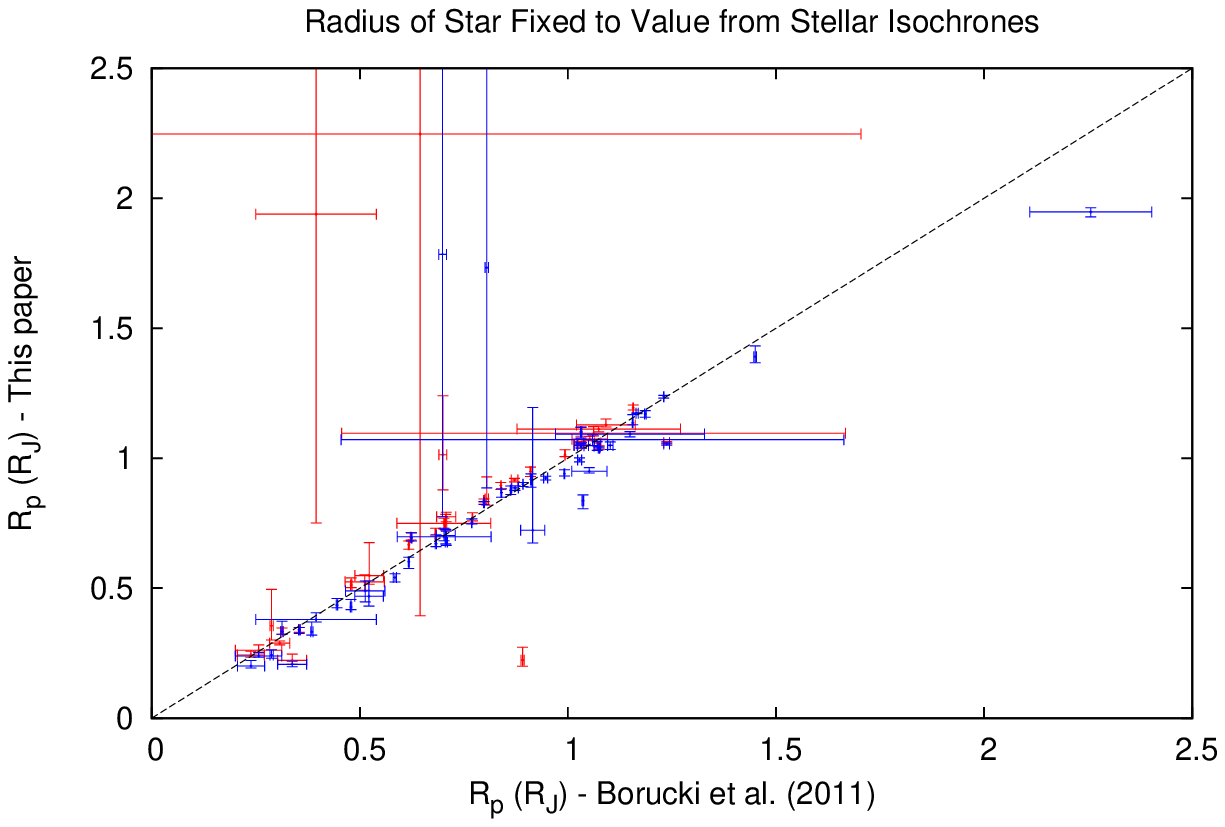}\\
\end{tabular}
\caption{Comparison of the values for the planetary radius as given by \citet{Borucki2011} and derived in this paper. Red and blue symbols correspond, respectively, to the results from the PDC and CLM light curves. The errorbars are computed assuming a fixed stellar radius, but taking into account the errors on the value of the ratio of the radii. In the left panel the stellar radius has been set to its value in the KIC, while in the right panel the stellar radius is computed via stellar isochrones from its given $T_{\star}$ value in the KIC. The dashed line delineates an expected 1:1 correlation.}
\label{borcompfig}
\end{figure*}

Given the stellar parameters and their associated errors, we proceeded with calculating physical parameters for each planet. From the orbital period of the system and the mass of the star, we calculated the semi-major axis of the planets via Newton's version of Kepler's Third Law. The radius of each planet, $R_{p}$, was calculated from the stellar radius and the value of the radius ratio derived in \S\ref{modelsec}. The obtained $R_{p}$ values are compared with those reported by \citet{Borucki2011} in Figure~\ref{borcompfig} by multiplying the derived value of the ratio of the radii from each study, including results from both the PDC and CLM data for ours, by the radius of the host star derived via both the KIC and stellar isochrones. Except for a handful of outliers, most values of the planetary radii derived via different parameter estimations seem to agree with the \citet{Borucki2011} results within $\sim$5\%. We note though that the radii of individual planet candidates can be significantly affected depending on whether their stellar radii are derived from the KIC or stellar isochrones, on average $\sim$20\%.

We also calculated the sub-stellar equilibrium temperature, maximum effective temperature, and brightness temperature of each planet following the same equations as in \citet{Cowan2011}, who themselves draw upon \citet{Hansen2008} and \citet{Burrows2008}. We calculated the equilibrium temperature of each planet at its sub-stellar point, $T_{0}$, using $T_{\star}$ and the semi-major axis of the system, $a$, via,

\begin{equation}
  T_{0} = T_{\star}\cdot(R_{\star}/a)^{0.5}
\end{equation}

\noindent We note that this expression assumes non-significant eccentricity effects on the heating of the planet by the star. 

We calculated the maximum effective temperature of the planet, $T_{\epsilon=0}$, using the equation

\begin{equation}
  T_{\epsilon=0} = (2/3)^{\frac{1}{4}}\cdot T_{0}
\end{equation}

\noindent assuming no albedo or heat recirculation.

Finally, to calculate the measured brightness temperature of each planet, $T_{b}$, we assume both the planet and star emit like blackbodies and compute $T_{b}$ by integrating their fluxes over the $Kepler$ passband from the equation

\begin{equation}
\label{Jeq}
  J = \frac{\int t_{\lambda}\cdot\lambda^{-5}\cdot(exp(\frac{hc}{\lambda kT_{b}})-1)^{-1} \cdot d\lambda}{\int t_{\lambda}\cdot\lambda^{-5}\cdot(exp(\frac{hc}{\lambda kT_{\star}})-1)^{-1} \cdot d\lambda}
\end{equation}

\noindent where $\lambda$ is a given wavelength, $t_{\lambda}$ is the net transmission of the telescope and detector at a given wavelength, $h$ is Planck's constant, $c$ is the speed of light, and $k$ is Boltzmann's constant.

In the cases where significant sinusoidal variations were detected in the light curves, i.e., cases where we found a positive value for $A_{L_{p}}$ with a detection level of at least 2$\sigma$, we treated the effect as real and accounted for it in the determination of the day-side flux of the planet by multiplying the values of $J$ in Eq.~\ref{Jeq} by (1+$A_{L_{p}}$). The only systems where significant sinusoidal variations were detected are KOI 2.01 and KOI 13.01, for both the PDC and CLM data, and both fixing eccentricity to zero and letting it vary. For KOI 1541.01 a significant sinusoidal variation was also found, but only for the CLM data and when allowing eccentricity to vary. Notice that, if real, the observed amplitude of sinusoidal variations can be either due to a significant albedo, and thus a varying amount of reflected light with phase, a significant temperature difference between the day and night sides of the planets, i.e., very little heat redistribution, and therefore varying amounts of emitted light with phase, or photometric beaming. It is possible that sinusoidal systematic signals could mask as significant phase variations, but in order to be detected as such by the residual-permutation error analysis we employed, the systematic feature would have to have a stable amplitude and period over the course of the 90-day observations, and have the period and phase of maximum amplitude coincide with the orbital period and secondary eclipse phase of the planet, which we deem unlikely.

In the above temperature calculations we have assumed no albedo and therefore all the observed planetary fluxes are due to thermal emission. However, the atmosphere of the planets can contain clouds or hazes which would reflect at least part of the incident stellar light. To account for those effects, we can estimate the different contribution amounts of reflected light to the measured planet-to-star surface brightness ratio, as 

\begin{equation}
\label{albeq}
  F_{a} = \frac{A \cdot R_{\star}^{2}}{a^{2}}
\end{equation}

\noindent where $A$ is the geometric albedo of the planet in the integrated $Kepler$ passband. Assuming different values of $A$ between 0.0 (no albedo) and 1.0 (purely reflective atmosphere), we can subtract the resultant value of $F_a$ from $J$ in order to remove the reflected light contribution from the measurements of the eclipse depths before computing the $T_{b}$ of the planet that accounts for the remaining, thermally emitted, light. Furthermore, given the measured surface brightness ratio and Equation~\ref{albeq}, we can determine the maximum possible geometric albedo of the planet in the $Kepler$ wavelength range, $A_{max}$, by assuming that all of the detected emission is solely due to reflected light. Setting $F_{a}$ = $J$ and solving for $A$, we obtain the expression

\begin{equation}
\label{maxalbeq}
 A_{max} = \frac{a^{2}J}{R_{\star}^{2}}
\end{equation}

Finally, we have computed robust errors for all the derived quantities, i.e.,  $T_0$, $T_{\epsilon = 0}$, $T_b$, $a$, $R_p$, $A_{max}$, and also $T_b / T_0$ (see next section) by re-calculating all their values 10,000 times, each time adding random Gaussian noise with amplitudes equal to the errors of the underlying quantities $J$, $P$, $k$, $T_*$, $M_*$, and $R_*$. The resulting median values of each parameter and their asymmetric Gaussian 1$\sigma$ errors are listed in Table~\ref{tab3} along with the detection significance of the secondary eclipse, $\sigma_{sec}$, and the luminosity ratio of the system, $L_{r}$ = $L_{p}$/$L_{\star}$, for both the PDC and CLM light curves, both letting eccentricity vary and fixing it to zero, and both using the stellar parameters derived from the KIC and via stellar isochrones. Negative values of $\sigma_{sec}$ mean that a negative value of $J$ was found, i.e., an increase of light at secondary eclipse, instead of the expected decrease. We deem those results unphysical, but note we can still use them to establish upper limits for the depth of the eclipse, and estimate the fraction of spurious eclipse detections in our analysis, as already described at the end of \S\ref{modelsec}.

\section{Statistical Properties of the Secondary Eclipse Emissions}
\label{trendssec}

\begin{figure*}
\centering
\begin{tabular}{cc}
   \epsfig{width=0.5\linewidth,file=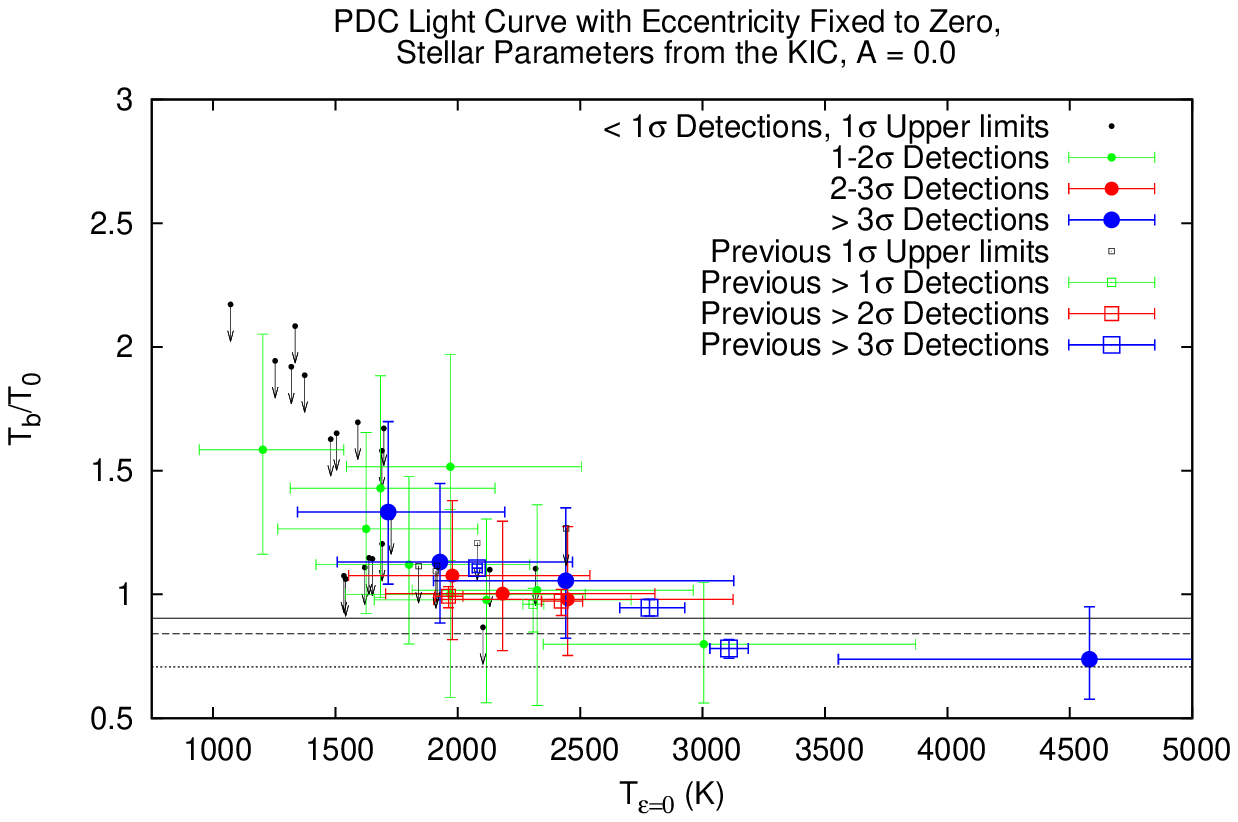} &
   \epsfig{width=0.5\linewidth,file=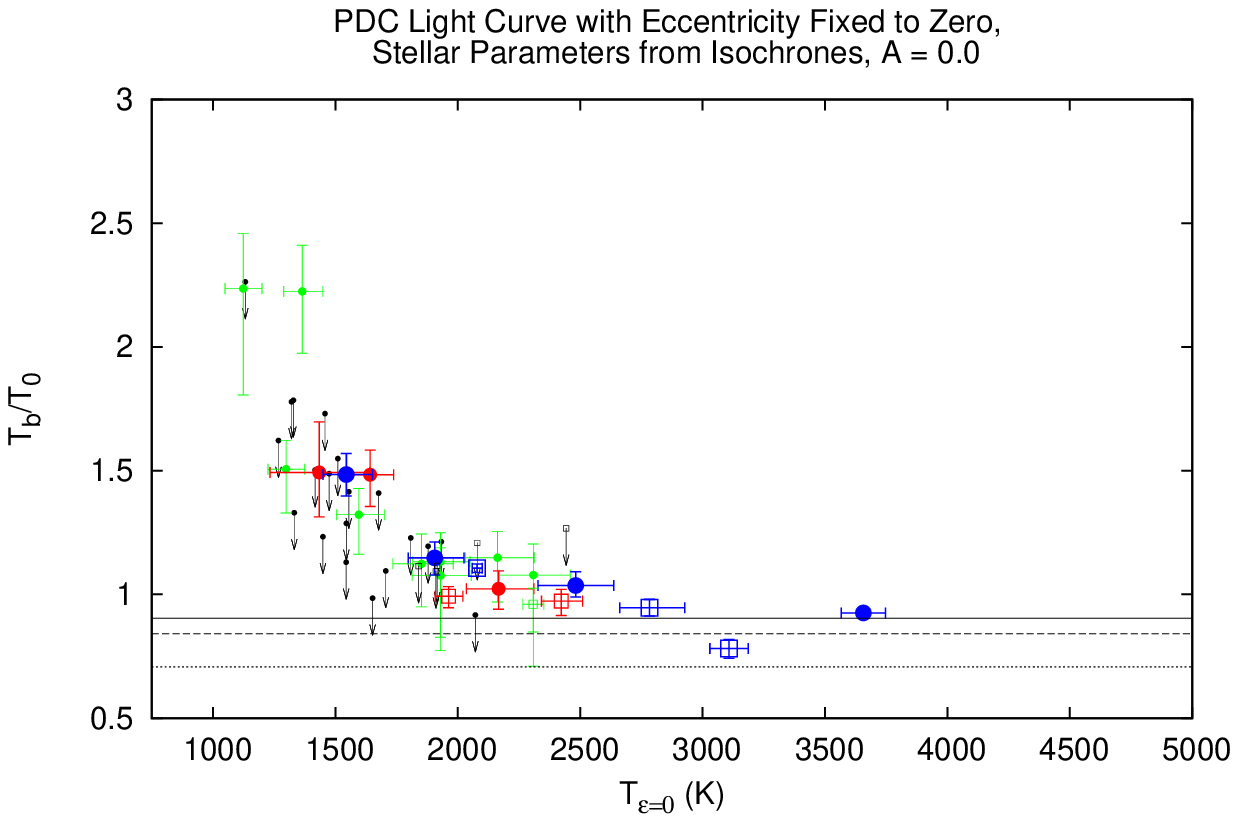} \\
   \epsfig{width=0.5\linewidth,file=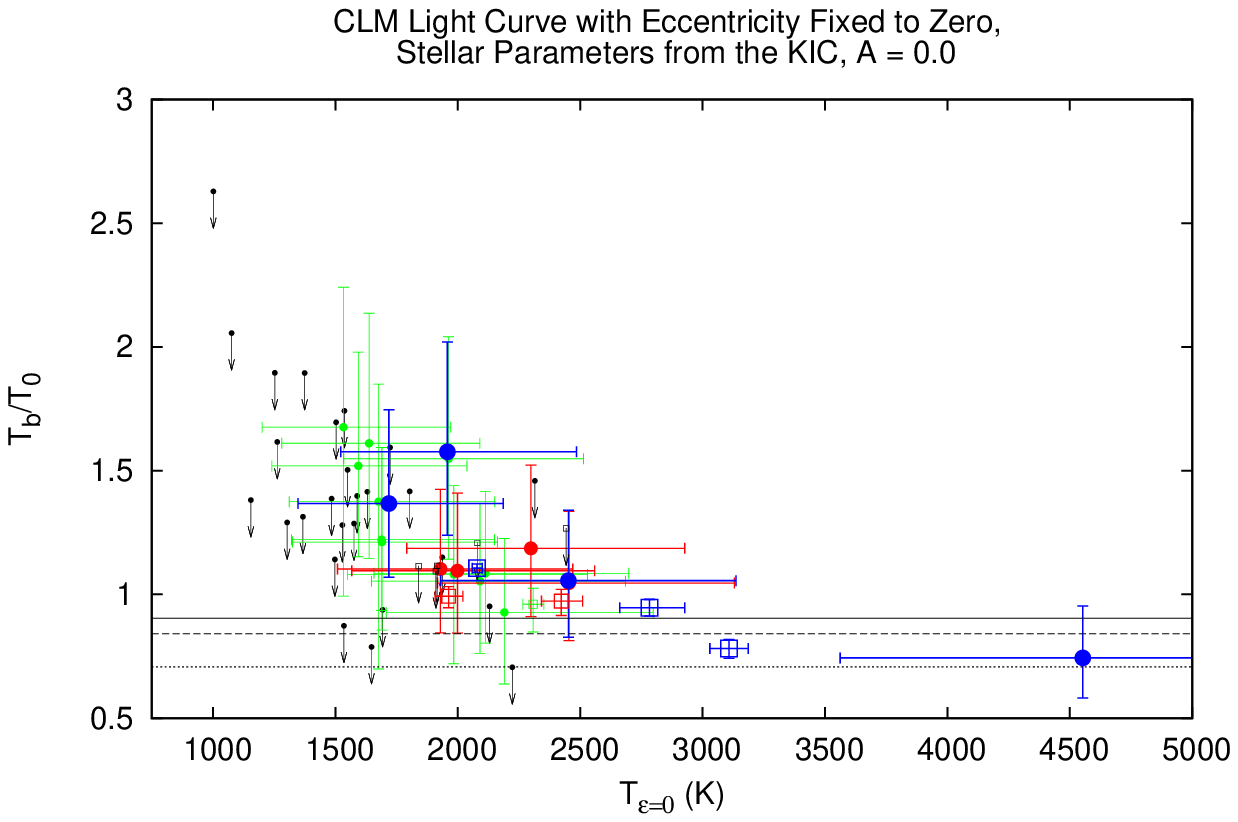} &
   \epsfig{width=0.5\linewidth,file=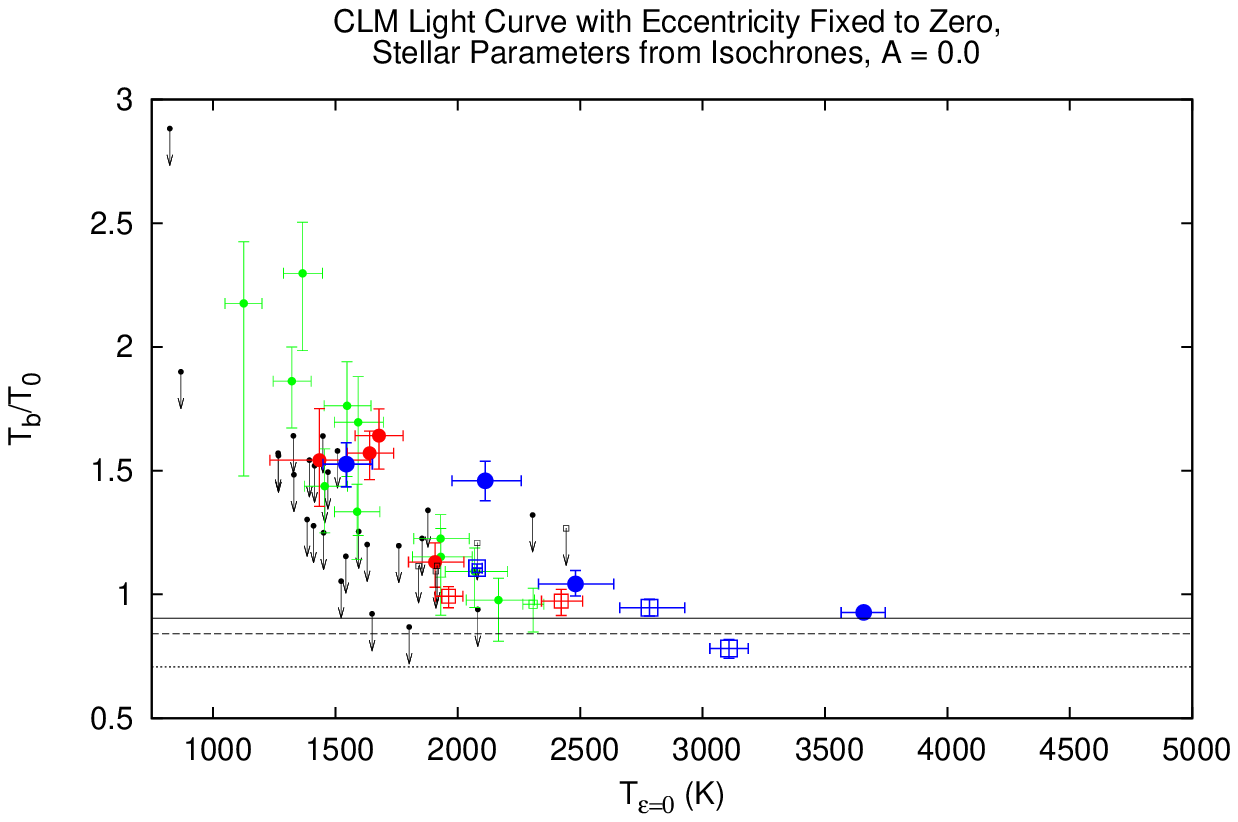} \\
\end{tabular}
\caption{Plots of the effective day side temperature ratio versus the maximum effective day side temperature when fixing eccentricity to zero. The values obtained when deriving stellar parameters from the KIC are shown in the left column, while values obtained when deriving stellar parameters from stellar isochrones are shown in the right column. Values obtained when using the $Kepler$ PDC light curves are shown in the first row, while values obtained when using the CLM pipeline are shown in the second row. Solid circles correspond to $Kepler$ systems modeled in this paper, while open squares are previously published detections or upper limits of exoplanet secondary eclipses at optical wavelengths. All errors are 1$\sigma$. The x-axis errorbars are not shown for the $<$1$\sigma$ detections for clarity. The solid, dashed, and dotted black lines in each figure correspond to the expected temperature ratio assuming no heat recirculation, a uniform day-side temperature, and a uniform planetary temperature respectively.}
\label{trfig}
\end{figure*}

\begin{figure*}
\centering
\begin{tabular}{cc}
   \epsfig{width=0.5\linewidth,file=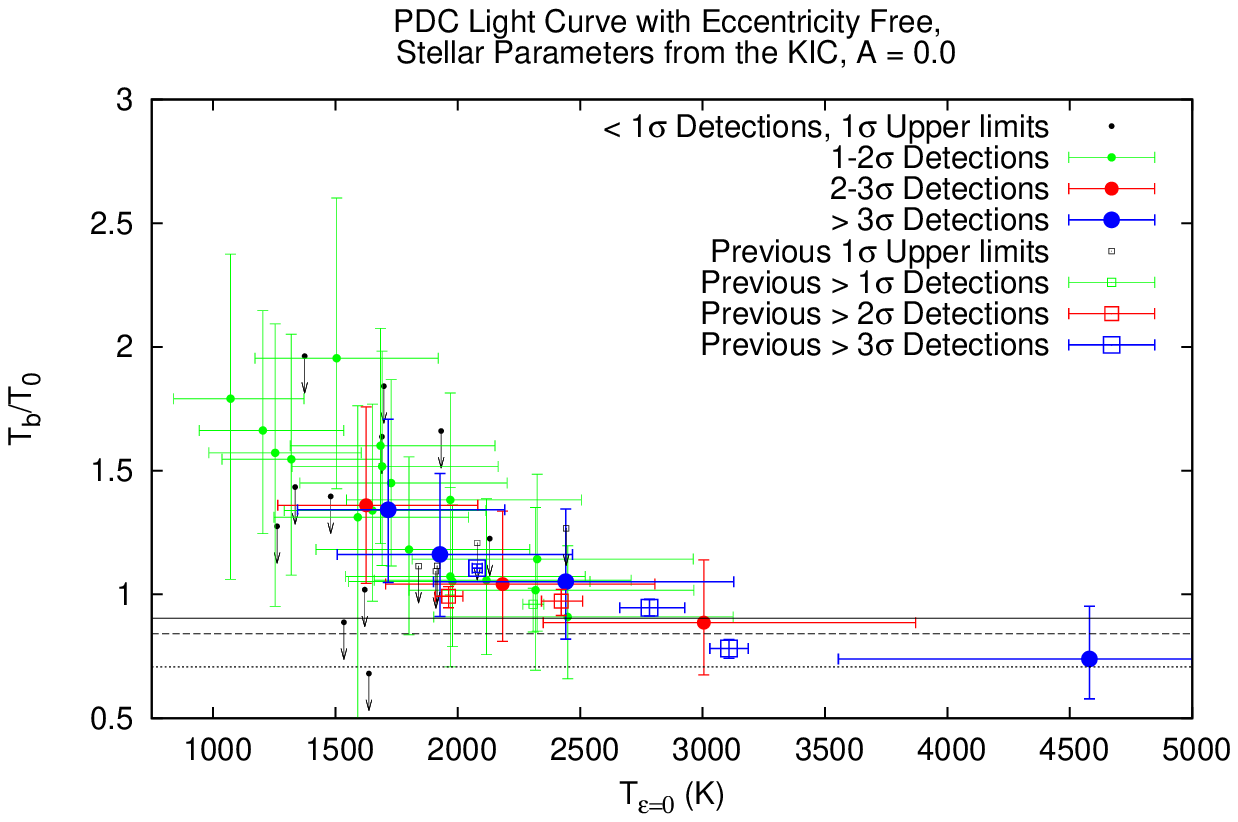} &
   \epsfig{width=0.5\linewidth,file=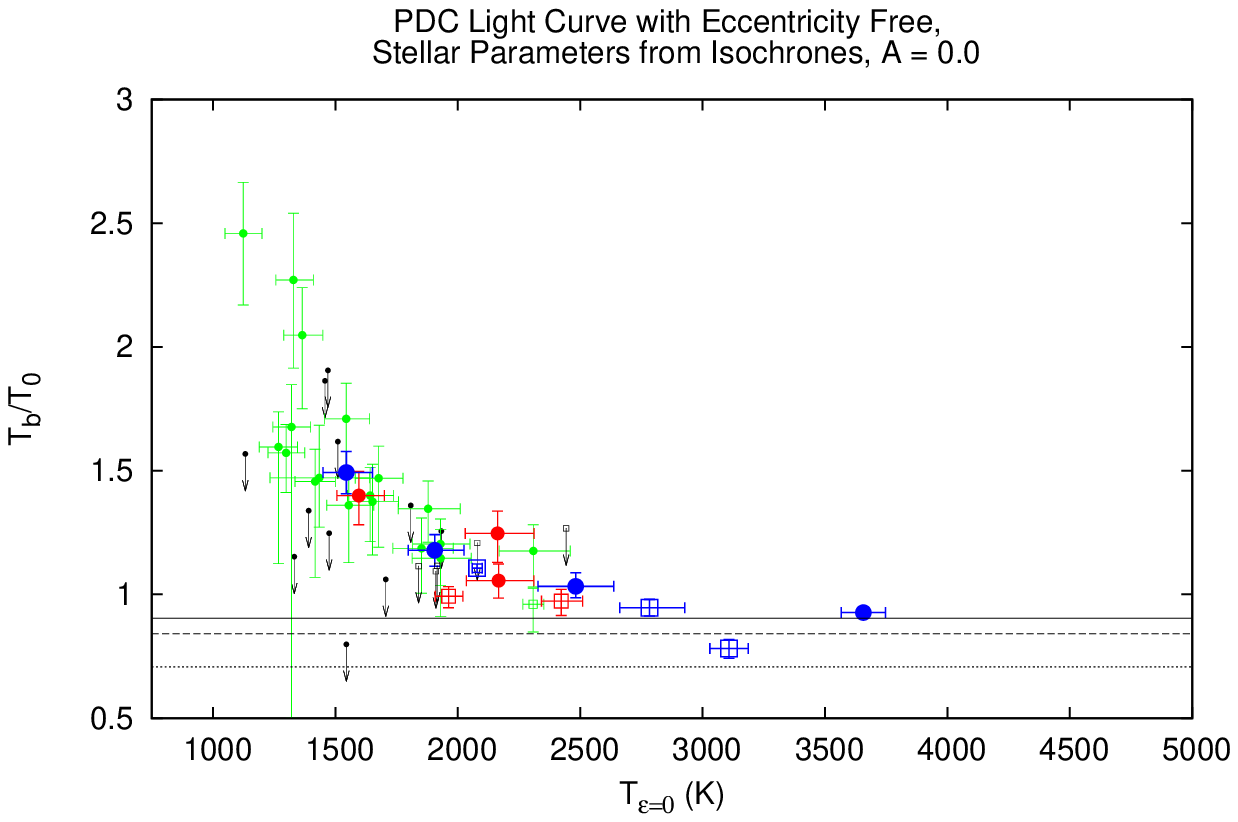} \\
   \epsfig{width=0.5\linewidth,file=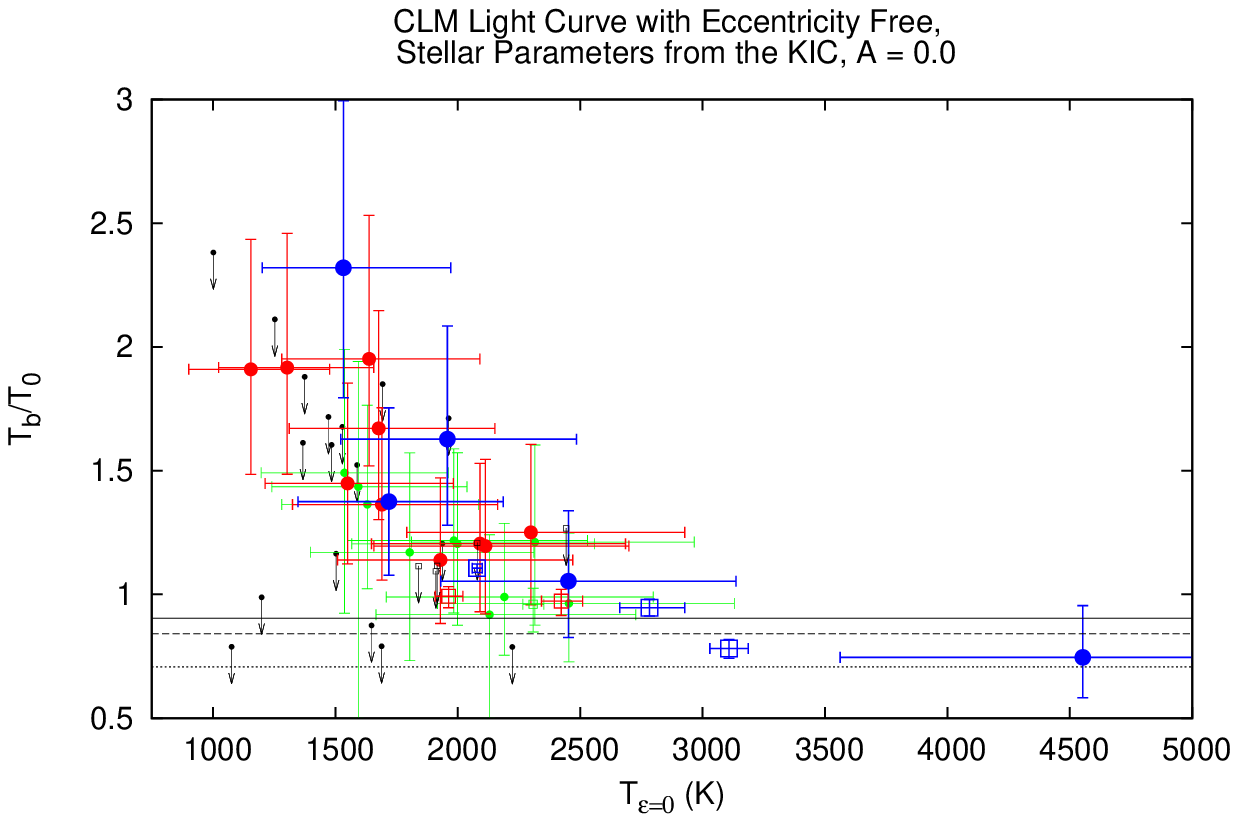} &
   \epsfig{width=0.5\linewidth,file=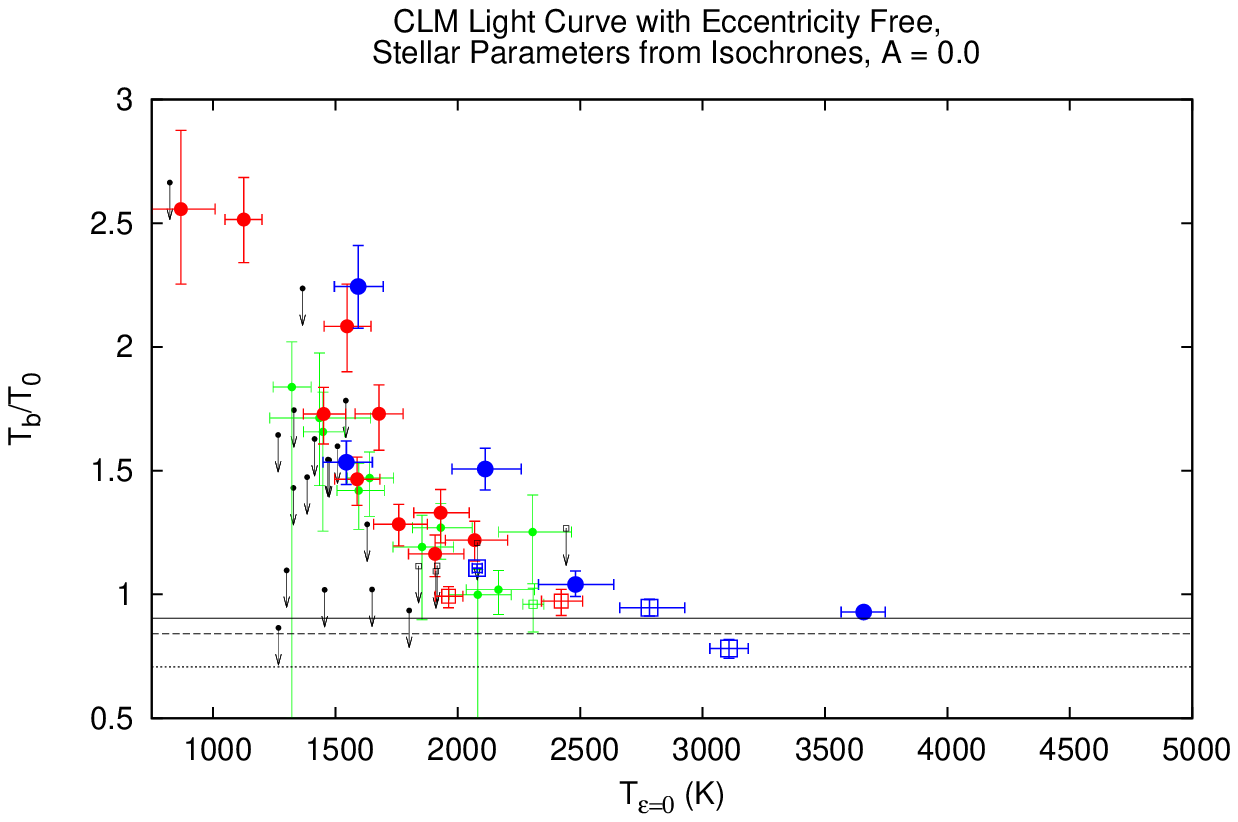} \\
\end{tabular}
\caption{Similar to Figure~\ref{trfig}, but with eccentricity allowed to vary.}
\label{trfig2}
\end{figure*}

Following \citet{Cowan2011}, we plot in Figure~\ref{trfig} the dimensionless $Kepler$ passband day-side brightness temperature ratio of each planet candidate in our sample, $T_{b}$/$T_{0}$, versus their maximum expected day-side temperature, $T_{\epsilon = 0}$, for the case of eccentricity fixed to zero. The different panels in the figure correspond to the results from the PDC and the CLM light curves, and both using the stellar parameters derived from the KIC and via stellar isochrones. In all the panels we have assumed zero albedo, which is equivalent to assuming that all the emission from the planet is thermal. (The case of non-zero albedos is considered below.) The 1-2$\sigma$, 2-3$\sigma$, and $>$3$\sigma$ detections, and 1$\sigma$ upper limits of the $<$1$\sigma$ detections, are represented by solid circles of different colors and sizes. In addition, the open squares correspond to planets with previously published secondary eclipse detections in the optical, i.e., CoRoT-1b \citep{Alonso2009b}, CoRoT-2b \citep{Alonso2009a}, Hat-P-7b \citep{Welsh2010}, Kepler-5b \citep{KippingBakos2011a}, Kepler-7b \citep{KippingBakos2011a,Demory2011}, and OGLE-TR-56b \citep{Sing2009,Adams2011}, as well as previously published secondary eclipse upper limits, i.e., HD209458 \citep{Rowe2008}, TrES-2b \citep{KippingBakos2011b}, and Kepler-4b, Kepler-6b, and Kepler-8b \citep{KippingBakos2011a}. Each point is shown with its 1$\sigma$ x and y-axis errorbars, except for the $<$ 1$\sigma$ detection upper limits, where the x-axis errorbars are omitted for clarity. Finally, the three horizontal lines in each plot indicate the expected values of $T_{b}$/$T_{0}$ for no energy redistribution, i.e., $f = 2/3$, a uniform day-side temperature, i.e., $f = 1/2$, and a uniform planetary temperature, i.e., $f = 1/4$ \citep[see][]{LopezMorales2007,Cowan2011}. In Figure~\ref{trfig2} we reproduce the same plots but with eccentricity allowed to vary.

Unless there is some extra emission at optical wavelengths that is not being accounted for, all the planets should lie below the $f = 2/3$ lines in Figures~\ref{trfig} and \ref{trfig2}. However, it is immediately apparent that the vast majority of candidates lie above that line. In addition, there appears to be a trend of increasing  $T_b / T_0$ with decreasing $T_{\epsilon = 0}$, with some possible detections approaching 2.5 times the maximum expected brightness temperature, i.e., nearly 40 times more flux than expected. All the planets with previously published secondaries also appear to follow the same trend, although they all have $T_{\epsilon = 0}$ $\gtrsim$ 2000 K. This trend will be discussed in more detail below.

We have explored several possible explanations for such large observed emissions at visible wavelengths: 1) A bias in the determination of stellar and planetary parameters or the secondary eclipse detection efficiency, 2) high albedos, which would make reflected light a major contributor to the planetary emission, 3) very large amounts of non-LTE or other thermal emission at optical wavelengths, 4) the presence of a significant source of internal energy generation within the planet, and 5) some of the candidates are in fact very low mass stars or brown dwarf companions, or background eclipsing binary blends.

{\it Potential~Biases}: The determined stellar parameters of $T_{\star}$, $M_{\star}$, and $R_{\star}$ can have significant uncertainties, as noted in \S\ref{paramsec}, and indeed can vary notably when taken from the KIC or computed from stellar isochrones based on $T_{\star}$. The values of those parameters are intimately tied to the derivation of the planetary parameters $T_{\epsilon = 0}$, $T_{0}$, $T_{b}$, $a$, $R_{p}$, $F_{a}$, and $A_{max}$, and that must be kept in mind when interpreting any possible results. For example, the stellar isochrones assume the stars are main-sequence, and thus if a host star is really a sub-giant or otherwise evolved, the stellar flux at the planet's surface would be underestimated. This would in turn cause an underestimated value of both $T_{\epsilon = 0}$ and $T_0$, and thus potentially overestimated values of $T_b / T_0$ at lower $T_{\epsilon = 0}$. However, it is unlikely that a large fraction of the examined systems are significantly evolved, and sub-giants would likely show telltale variations in their light curves. Given this, and that as far as we know there is no other preferential bias in the determination of the stellar parameters, we would not expect this problem to systematically influence the results presented in Figures~\ref{trfig} and \ref{trfig2}.

When examining secondary detection efficiency, we note that the derived upper limits on the secondary eclipse depths are roughly at the same level as the noise of the $Kepler$ data. That level of noise is consistent among the set of $Kepler$ light curves we have examined. Thus, as we search the data for planets with lower values of $T_{\epsilon =0}$ and $T_{0}$, the corresponding upper limit for $T_{b}$/$T_{0}$ naturally increases, as the expected eclipse depth decreases while the noise level remains the same. This introduces an artificial trend of higher  $T_{b}$/$T_{0}$ upper limits as $T_{\epsilon =0}$ decreases, which can be seen in Figures~\ref{trfig} and \ref{trfig2}. Similarly, when examining the $\sim$1$\sigma$ detections, about $\sim$32\% of the detections will be statistically spurious, as exemplified in the data and already discussed in \S\ref{modelsec}. We thus expect an artificial trend of higher $T_{b}$/$T_{0}$ values with decreasing values of $T_{\epsilon = 0}$ for $\sim$32\% of the $\sim$1$\sigma$ detections. However, when we move to the $\gtrsim$ 2$\sigma$ detections, the expected rate of spurious detections drop to $\lesssim$ 5-10\%. The data in Figures~\ref{trfig} and \ref{trfig2} still reveal a significant trend of increasing $T_{b}$/$T_{0}$ values with decreasing $T_{\epsilon = 0}$ in the $\gtrsim$~2$\sigma$ detections, (including as well previous secondary eclipse detections from the literature), so we conclude this trend is real and due to either increasing albedos or emission features in the visible as the atmospheric temperature of the planets decrease. The hypothesis of high albedos is further discussed below. As for emission features, a literature search on this topic does not reveal any known physical processes that would produce this effect, so further theoretical work might be necessary.

\begin{figure*}
\centering
\begin{tabular}{cc}
  \epsfig{width=0.5\linewidth,file=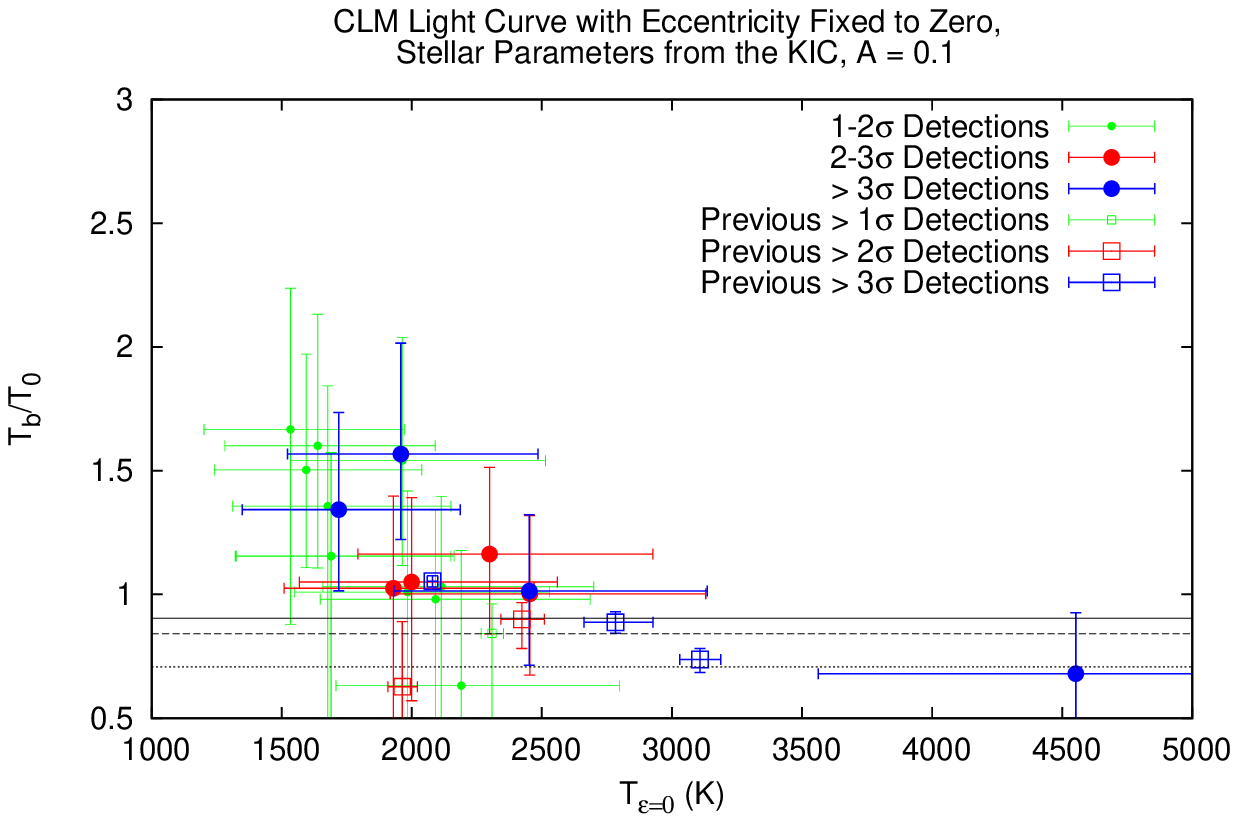} &
  \epsfig{width=0.5\linewidth,file=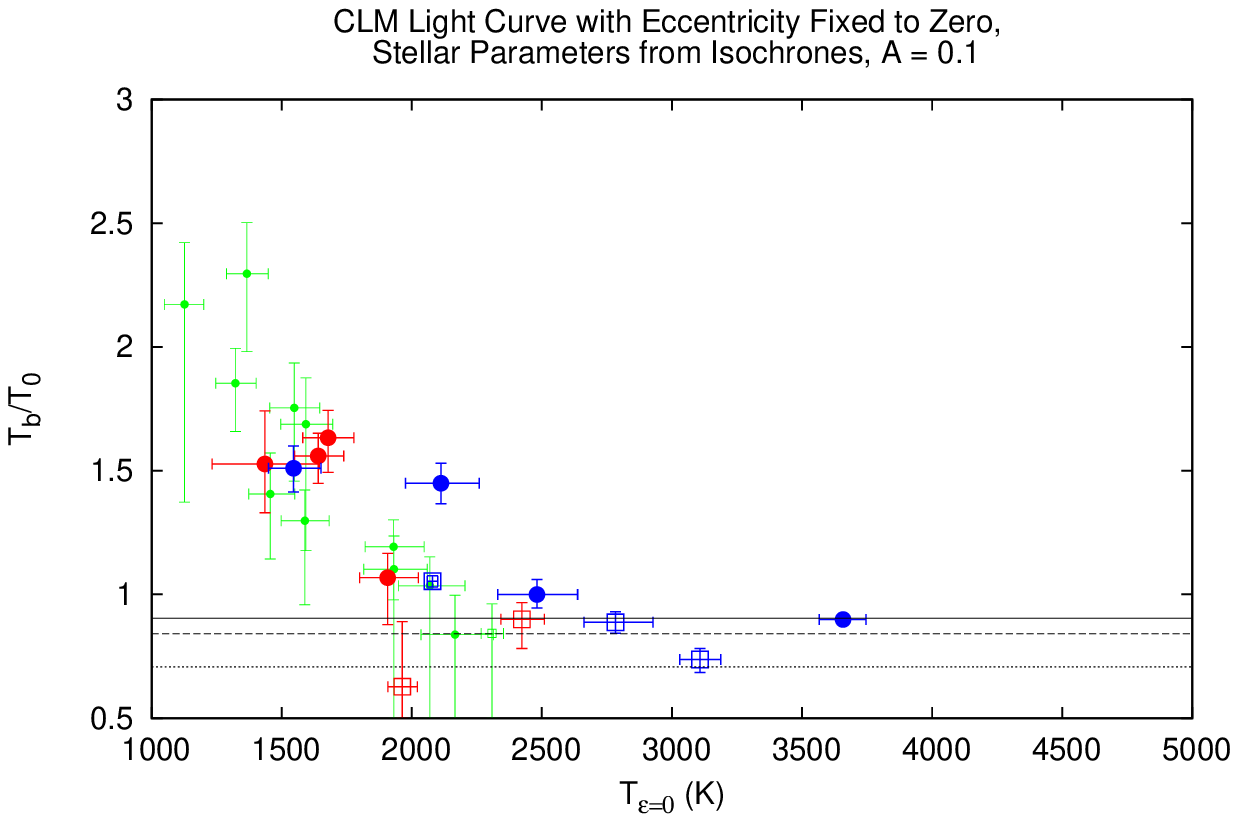} \\
  \epsfig{width=0.5\linewidth,file=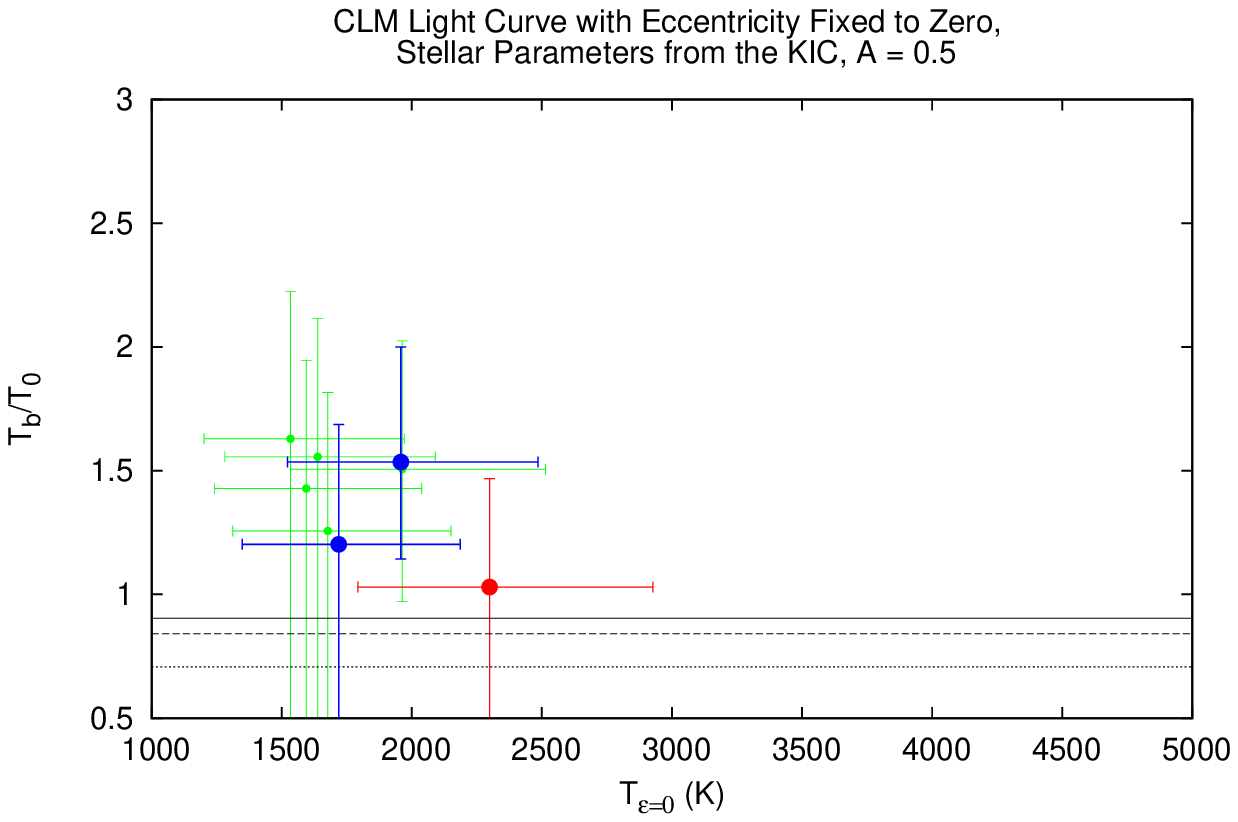} &
  \epsfig{width=0.5\linewidth,file=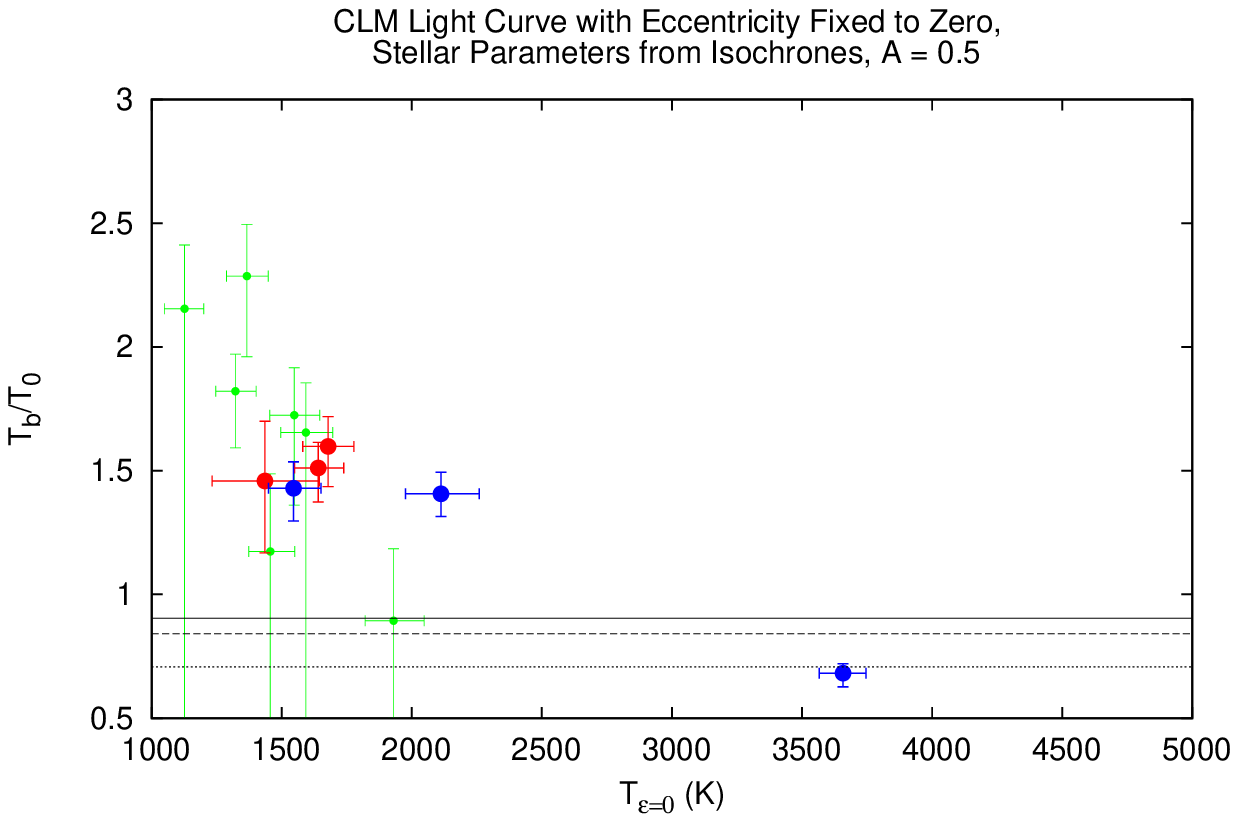} \\
  \epsfig{width=0.5\linewidth,file=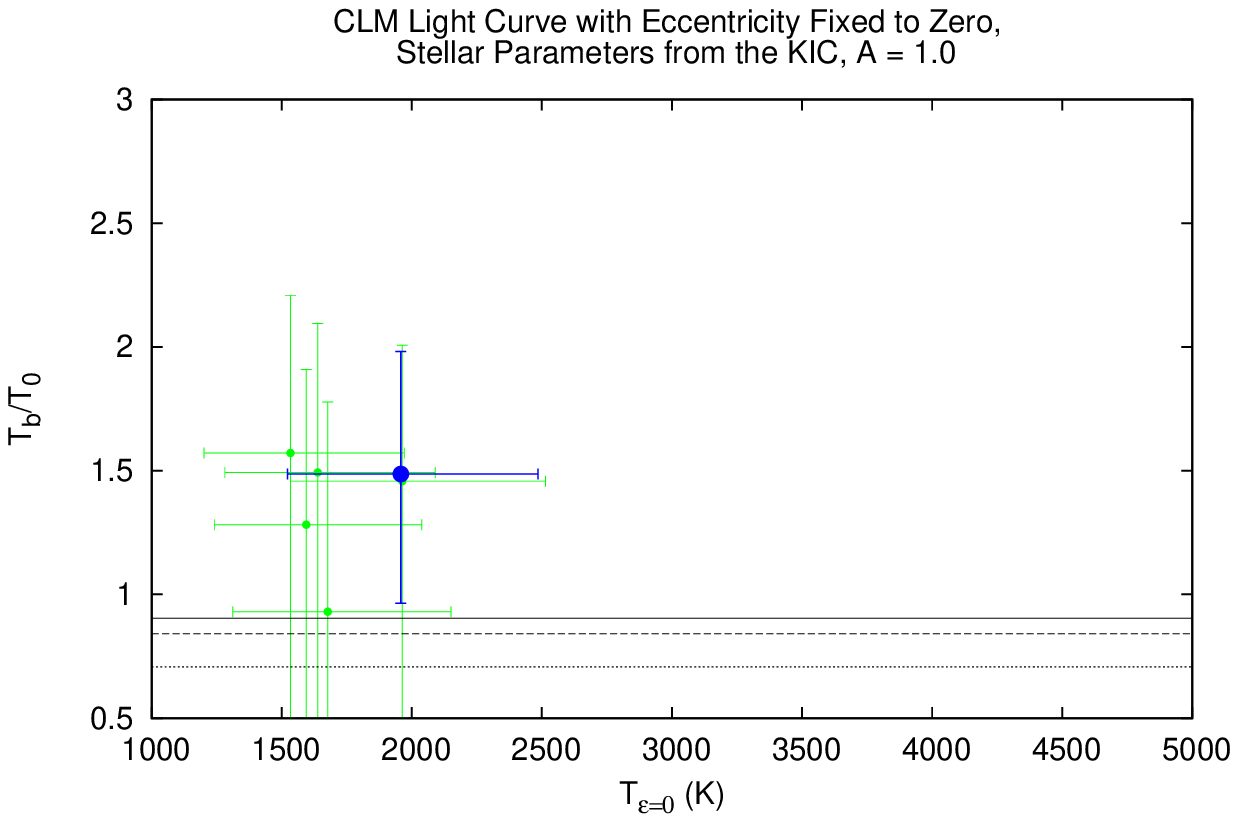} &
  \epsfig{width=0.5\linewidth,file=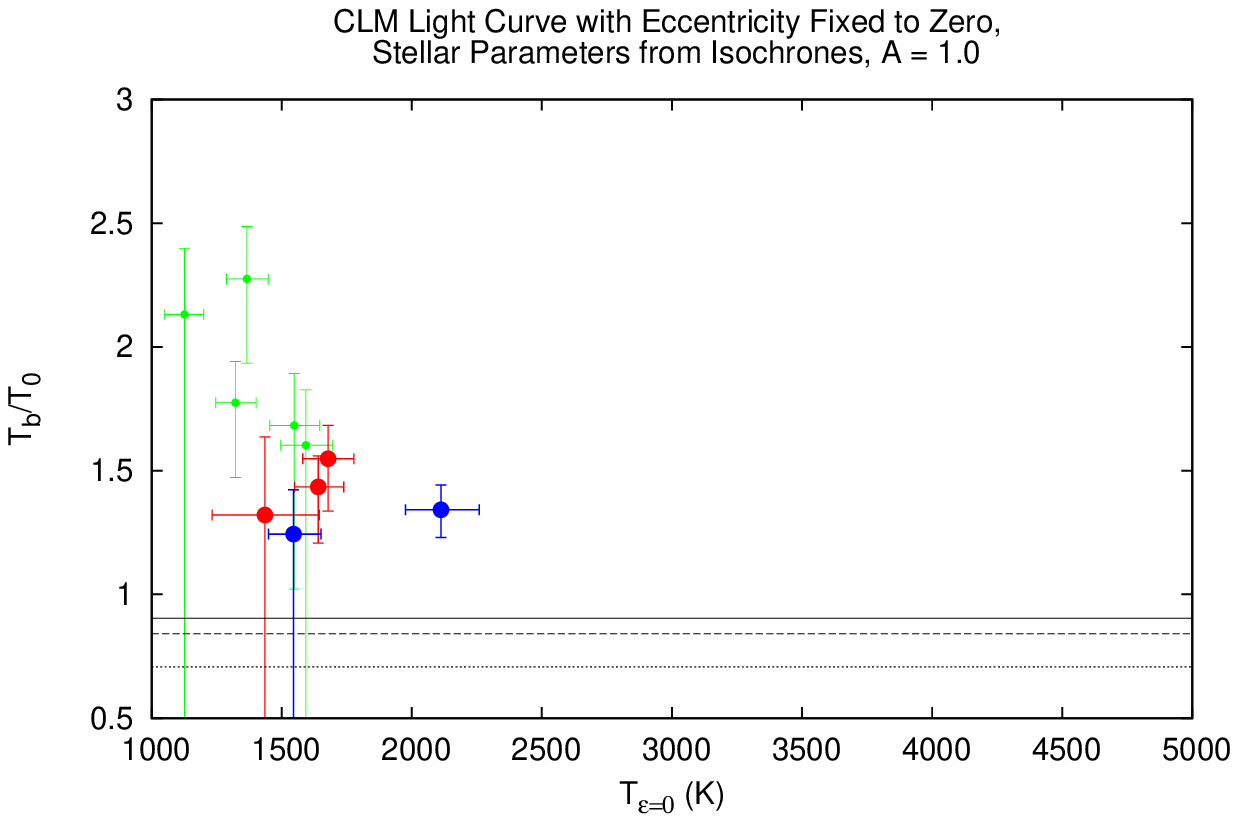} \\
\end{tabular}
\caption{Plots of the effective day side temperature ratio versus the maximum effective day side temperature for, from top to bottom, albedos of 0.1, 0.5, and 1.0, using the CLM pipeline light curves and assuming no eccentricity. The values obtained when deriving stellar parameters from the KIC are shown in the left column, while values obtained when deriving stellar parameters from stellar isochrones are shown in the right column. Solid circles correspond to $Kepler$ systems modeled in this paper, while open squares are previously published detections or upper limits of exoplanet secondary eclipses at optical wavelengths. The solid, dashed, and dotted black lines in each figure correspond to the expected temperature ratio assuming no recirculation, a uniform day-side temperature, and a uniform planetary temperature respectively. Systems that disappear from the plots when moving from low to high albedos can be fully explained by reflected light, while systems that remain at A = 1.0 still present excess emission that cannot be explained solely by reflected light.}
\label{albfig}
\end{figure*}

{\it High~Albedos}: As mentioned in the introduction, some recent studies suggest high albedos for some known hot giant planets \citep[e.g.][]{Berdyugina2011,KippingBakos2011a,Demory2011}. To examine the possibility that the excess flux observed in our list of secondary eclipse candidates is due to albedo, we have recomputed the expected normalized brightness temperature $T_{b}$/$T_{0}$ of each candidate when assuming increasing values of the albedo. The reflected light contribution is computed using equation~\ref{albeq} and subtracted from the total flux measured for each object. $T_{b}$ is then recomputed using the remaining flux, assuming it is solely due to thermal emission. The results are shown in Figure~\ref{albfig} for three different albedos, $A$ = 0.1, 0.5, and 1.0, using the secondary eclipse depths measured from the CLM pipeline light curves for $e = 0$, (the eclipse depths measured from the other light curves described in \S\ref{datasec} give similar results).

As expected, $T_{b}$/$T_{0}$ decreases as the albedo increases, and many of the points in Figure~\ref{albfig} go below the $T_{b}$/$T_{0}$ = (2/3)$^{1/4}$, i.e., $A$ =0 and $f$ = 2/3, limit once high enough albedos are assumed. The emission of 53\% of the planet candidates can be interpreted as a combination of reflected light and thermal emission when we assume a geometric albedo of $A$ = 0.5, set $e$ = 0, and derive stellar parameters from the KIC, though only 31\% when deriving stellar parameters from stellar isochrones. Those levels of reflectivity might indicate the presence of clouds, haze, or Rayleigh scattering in the atmosphere of those planets, with a general trend of increasing albedo with decreasing planetary temperature responsible for the trend of increasingly excess emission at lower planetary temperatures. We note, however, that a significant number of points, 40\% and 63\% when deriving stellar parameters from the KIC and stellar isochrones respectively, with $e$ = 0, still remain above $T_{b}$/$T_{0}$ = (2/3)$^{1/4}$, even if we assume perfectly reflecting planets, i.e. $A$ = 1.0. Many of these remaining systems are at low values of $T_{\epsilon = 0}$, and some of those points are $\gtrsim$ 2$\sigma$ detections, so there is still excess emission and a correlation of $T_{b}$/$T_{0}$ with decreasing $T_{\epsilon = 0}$ that cannot be explained by reflected light, and needs to be explained in some other way.

{\it Non-LTE~or~Other~Thermal~Emission}: We have used the brightness temperature parameter $T_{b}$ to represent the amount of thermal emission from each planet, assuming that the planets emit as blackbodies. In that case, if the emission of the planet yields a $T_{b}$ larger than that predicted by $f$ = 2/3 and $A$ = 0, that emission is above the so-called equilibrium temperature. However, the atmospheres of exoplanets do not necessarily emit as blackbodies, and some spectral models of hot Jupiters \citep[e.g.,][]{Fortney2008} predict significantly higher emission levels in the optical region covered by the $Kepler$ bandpass ($\sim$ 0.4 - 0.9 $\mu m$) compared to blackbody approximations. The emission spectrum of a planet will depend strongly on its atmospheric composition and Temperature-Pressure (T-P) profile. Although some models including the presence of strong absorbers, such as TiO and VO, have been proposed \citep{Hubeny2003,Fortney2008,Burrows2008}, uncertainties in the T-P profiles and the lack of previous observational data limit the reliability of those models. In addition, and as already mentioned above, there is currently no model that predicts the increase of $T_{b}$/$T_{0}$ for decreasing $T_{\epsilon = 0}$ observed in Figures~\ref{trfig} and \ref{trfig2}.

Another possibility is that the large amount of stellar irradiation these planets receive induces resonant and non-resonant fluorescent transitions by exciting the chemical species present in their upper atmospheres. Fluorescence has been measured in Solar System objects, i.e., Titan, Saturn and Jupiter, at IR, UV, and X-ray wavelengths \citep[e.g.,][] {YelleGrith2003,LopezValverde2005,Cravens2006,Lupu2011}, and possibly in the NIR for the exoplanet HD189733b \citep{Swain2010}, but we could not find any observations of or theoretical work on non-LTE/fluorescence emission in the 0.4 - 0.9$\mu m$ wavelength range covered by $Kepler$. However, it is expected that non-LTE emission lines in exoplanetary atmospheres will not be significantly broadened by collisions, and will appear instead as sharp emission features (Martin-Torres, priv. comm.). Considering the amount of energy required to produce these fluorescent emissions, it is unlikely that, given their narrow emission range, they would be luminous enough to significantly increase the measured emission levels over the very wide Kepler bandpass above the expected LTE emissions observed in Figures~\ref{trfig} and \ref{trfig2}.

{\it Significant~Internal~Energy~Generation}: Our own Jupiter radiates about 1.6 times as much energy as it receives from the Sun. The additional heat source is generally attributed to either residual heat left over from the initial Solar System nebula collapse, or ongoing slow contraction of the planet's core. However, if the planet candidates in our list were undergoing a similar internal energy generation process, $T_b / T_0$ would only reach up to about 1.6$^{1/4}$ = 1.12, not high enough to explain the emission of objects in our sample with $T_{\epsilon =0}$ $\lesssim$ 1500 K.

\begin{figure*}
\centering
\begin{tabular}{cc}
  \epsfig{width=0.5\linewidth,file=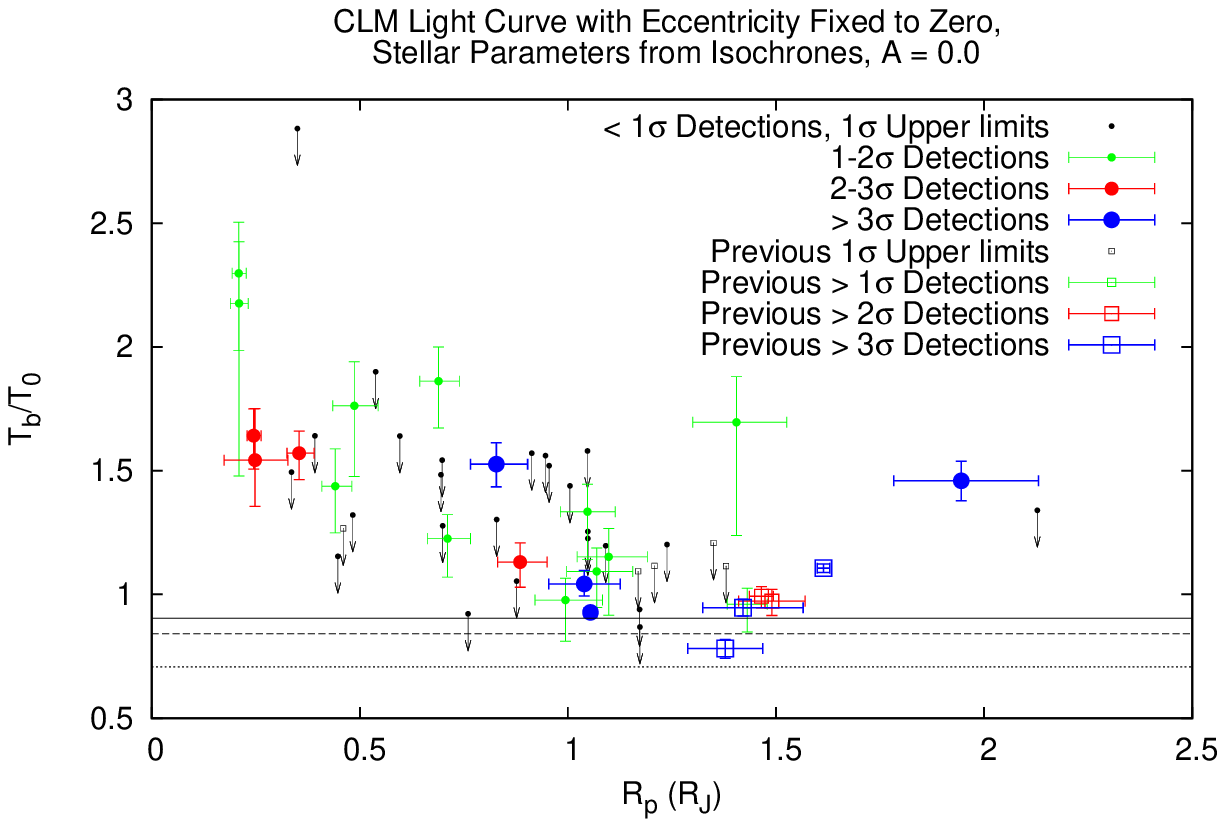}&
  \epsfig{width=0.5\linewidth,file=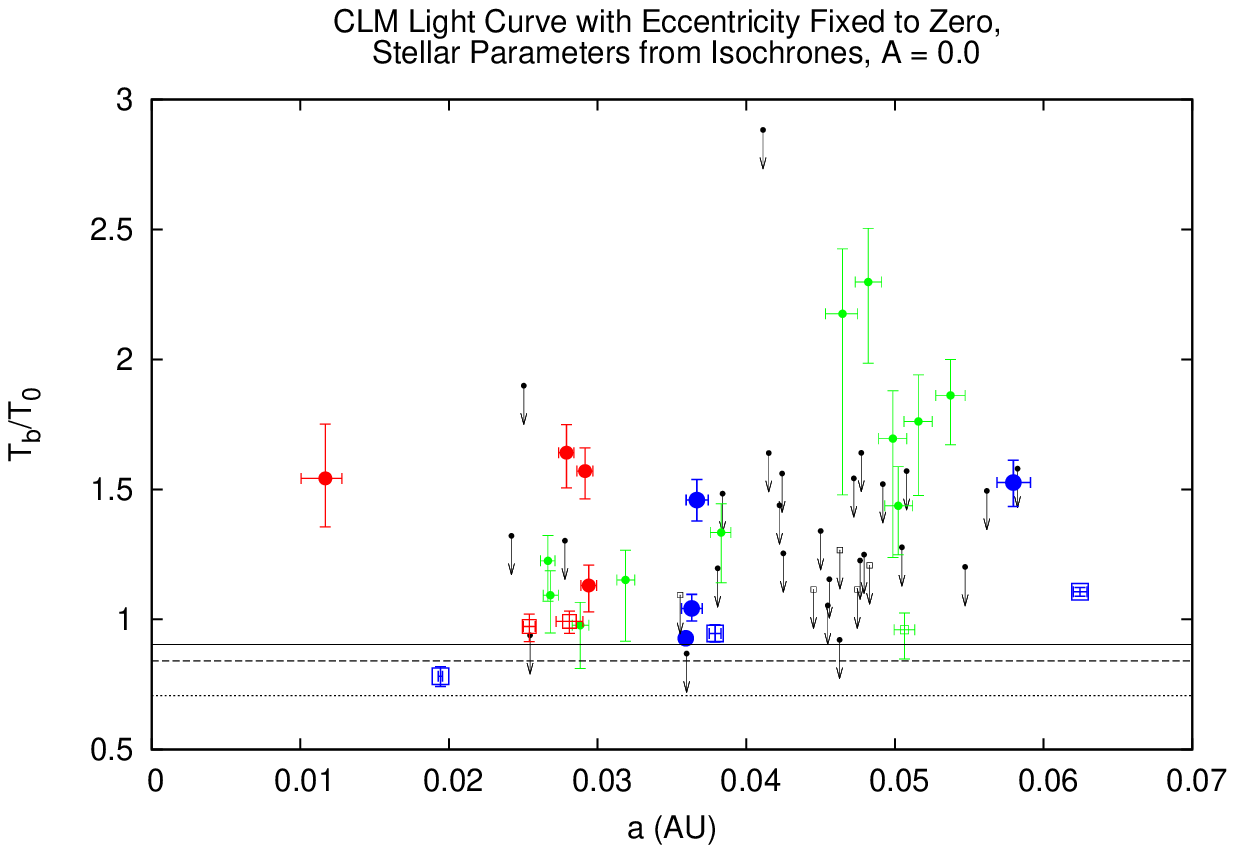}\\
  \epsfig{width=0.5\linewidth,file=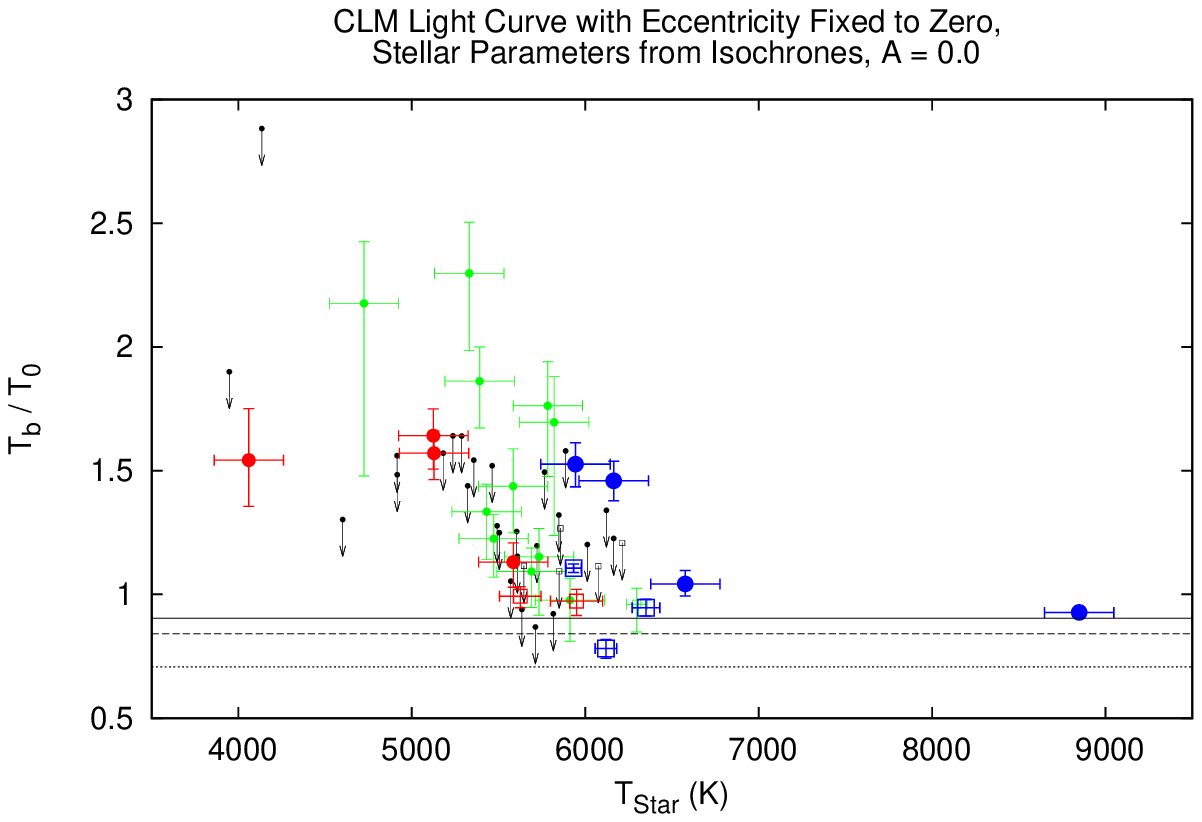}&
  \epsfig{width=0.5\linewidth,file=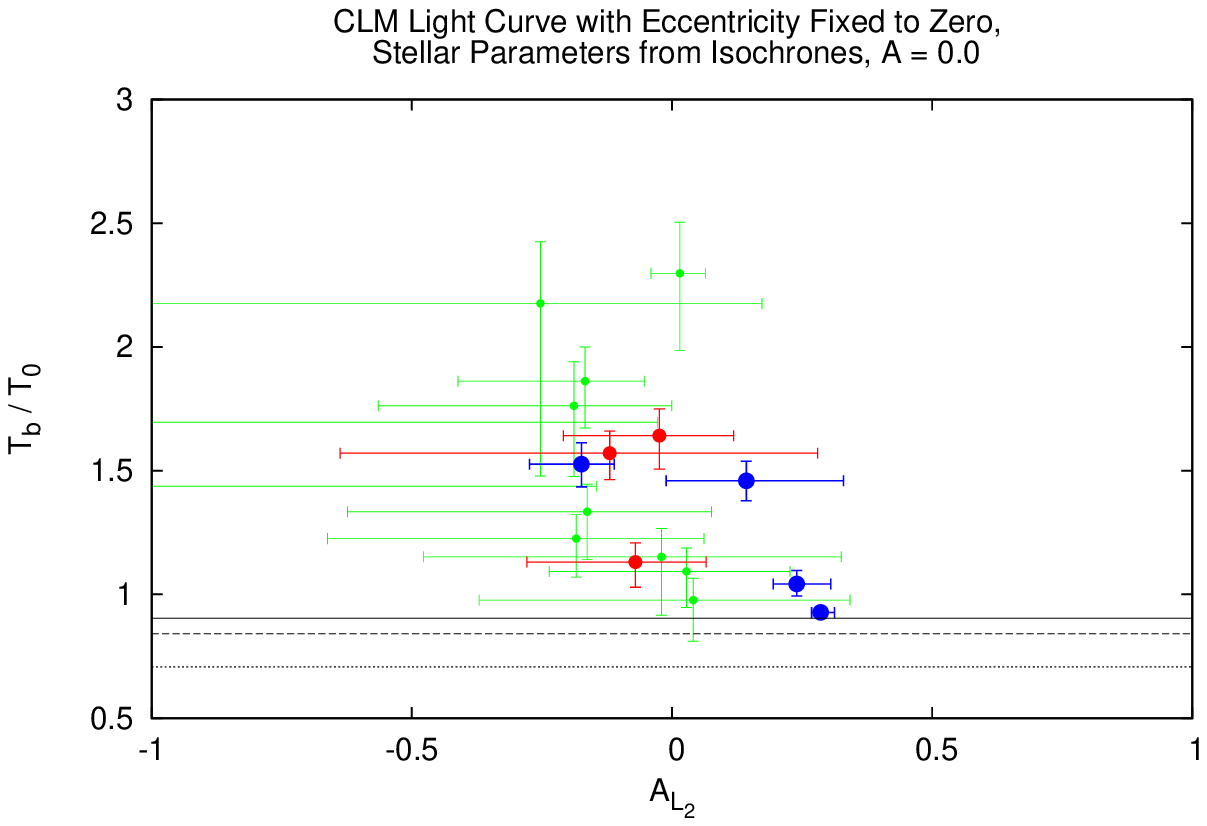}\\
\end{tabular}
\caption{Plots of the effective day side temperature ratio versus the radius of the planet (top-left), the semi-major axis of the system (top-right), the effective temperature of the star (bottom-left), and the amplitude of the sine curve applied to the planet's luminosity (bottom-right), using the CLM light curves, assuming no eccentricity, and deriving stellar parameters from stellar isochrones. Results for the PDC light curves are not shown, but produce similar results. Solid circles correspond to $Kepler$ systems modeled in this paper, while open squares are previously published detections or upper limits of exoplanet secondary eclipses at optical wavelengths. The x-axis errorbars are not shown for the $<$1$\sigma$ detections for clarity. The solid, dashed, and dotted black lines in each figure correspond to the expected temperature ratio assuming no recirculation, a uniform day-side temperature, and a uniform planetary temperature respectively.}
\label{rtrendfig}
\end{figure*}

{\it Low~Mass~Stars,~Brown~Dwarfs,~or~Blends}: The last possibility we consider is that some of the objects in the list are in fact brown-dwarfs or low-mass stars, but this assumption also posses some problems. Exoplanet search results over the years have revealed what appears to be a ``brown dwarf desert'' within orbital separations from the host star of $a$ $\lesssim$ 5 AU, for solar type stars \citep[see e.g.,][]{Grether2006}. However, the discovery of CoRoT-3b, a 21.7 $M_{Jup}$ object orbiting at a separation of only 0.057 AU around an F3-type star, has opened some debate about whether this object is really a brown dwarf, or if planets more massive than the defined Deuterium burning limit can form around stars more massive than the Sun. To test this idea we plot in Figure~\ref{rtrendfig} the measured $T_{b}$/$T_{0}$ of each planet candidate versus the effective temperature of the star, $T_{\star}$. We see, however, no clear correlation between the temperature of the host star and an excess brightness of the planet candidates, and an error-weighted linear fit does not yield a statistically significant slope. We also plot in Figure~\ref{rtrendfig} the values for $T_{b}$/$T_{0}$ versus $R_{p}$, $a$, and $A_{L_{p}}$, but do not see any significant linear correlations either.

\begin{figure*}
\centering
\begin{tabular}{cc}
  \epsfig{width=0.5\linewidth,file=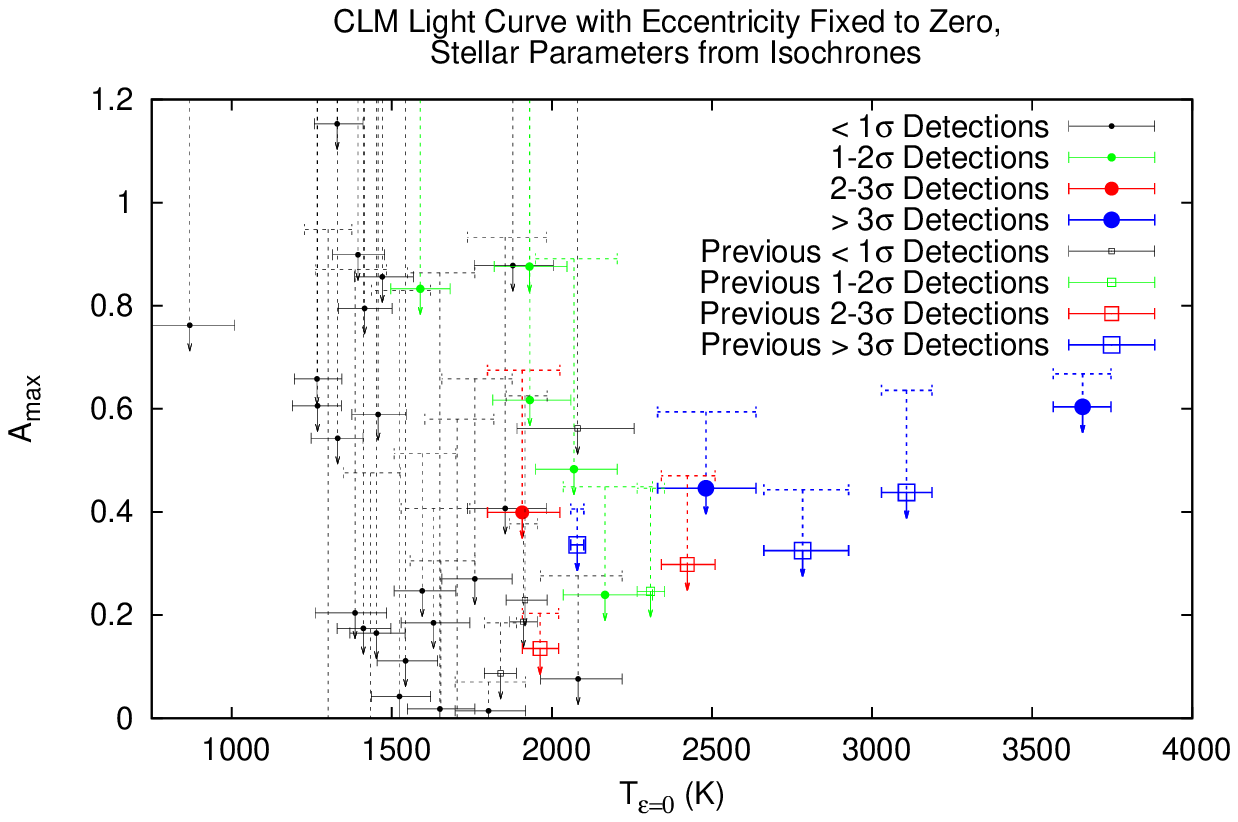}&
  \epsfig{width=0.5\linewidth,file=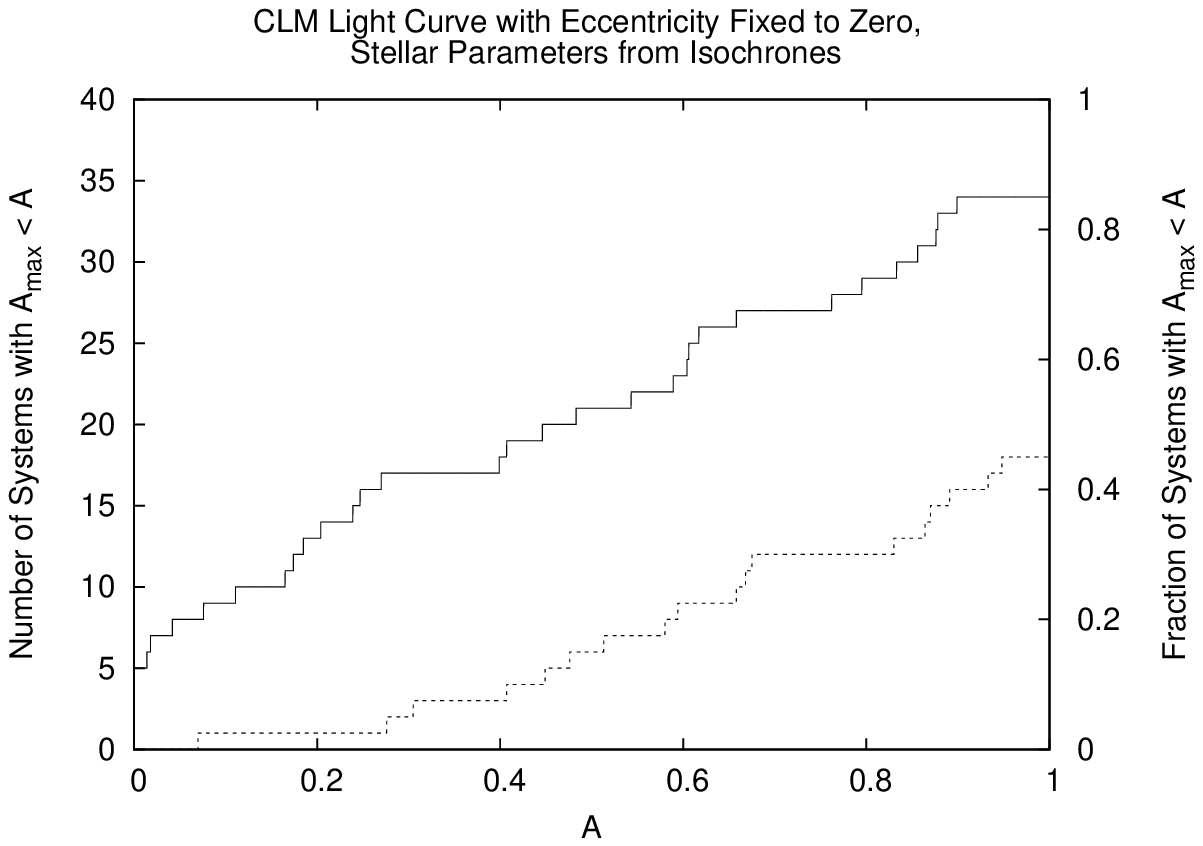}\\
  \epsfig{width=0.5\linewidth,file=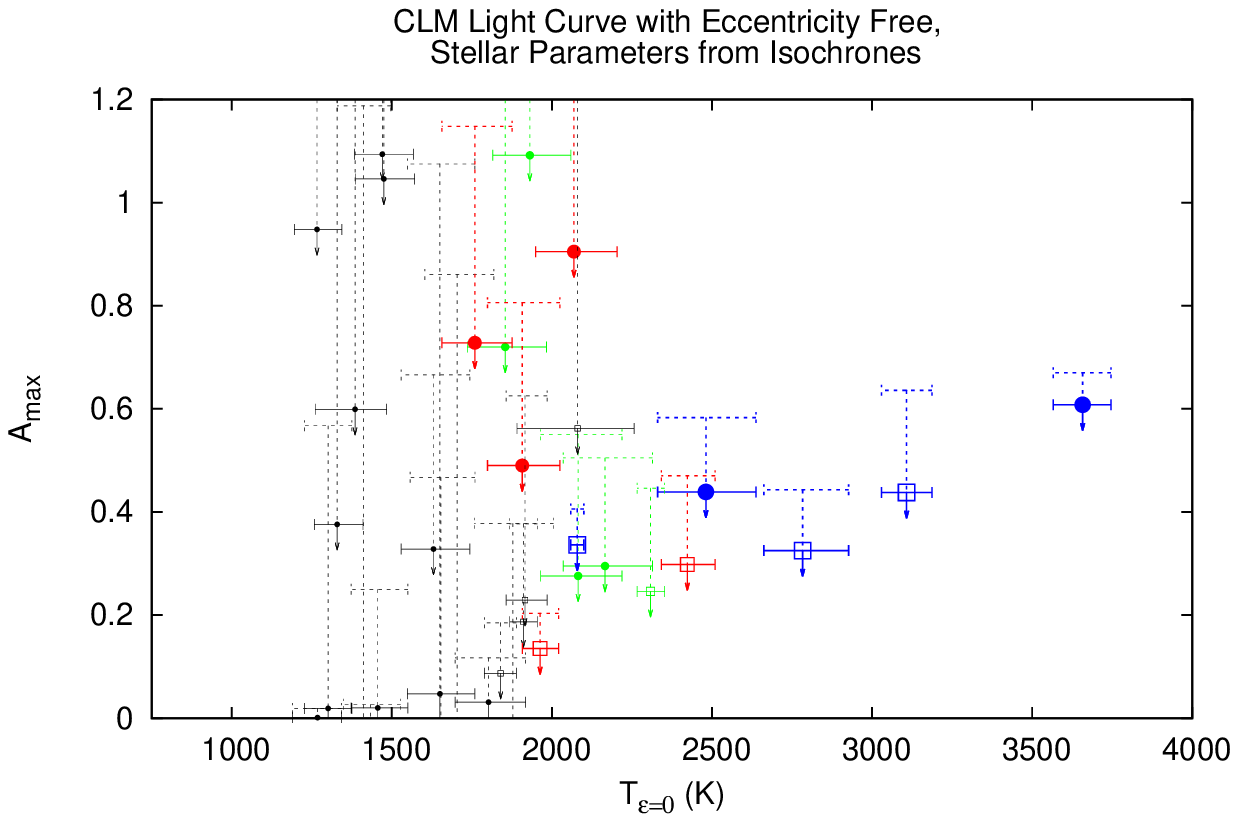}&
  \epsfig{width=0.5\linewidth,file=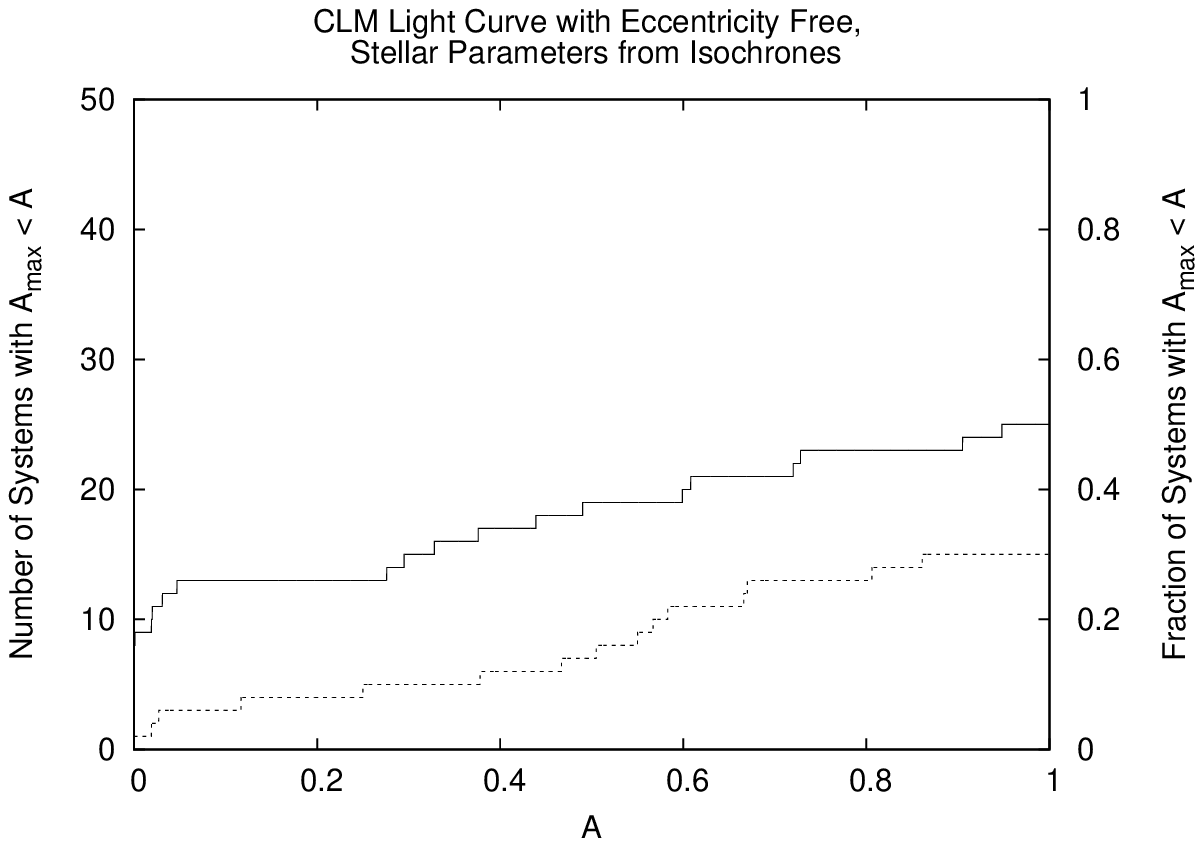}\\
\end{tabular}
\caption{Left Column: Plots of the 1$\sigma$ and 3$\sigma$ upper limits, delineated by solid and dashed lines respectively, on the maximum possible albedo values versus the maximum effective day side temperature when deriving stellar parameters from stellar isochrones, using the CLM pipeline light curves. Right Column: Plots of the cumulative number of systems and total fraction of all systems modeled in this paper that have their upper limit of $A_{max}$ below a given value of $A$, for both 1$\sigma$ and 3$\sigma$ upper limits. Values obtained when fixing eccentricity to zero are shown in first row, while values obtained when allowing eccentricity to vary are shown in the second row. Solid circles correspond to $Kepler$ systems modeled in this paper, while open squares are previously published detections or upper limits of exoplanet secondary eclipses at optical wavelengths. The results when using the $Kepler$ PDC curves are not plotted, but are very similar to the presented CLM light curves.}
\label{maxalbfig}
\end{figure*}

We also utilize the upper limits on possible secondary eclipses we derived to examine potential trends in $A_{max}$ as computed using Eq.~\ref{maxalbeq}. In Figure~\ref{maxalbfig} we plot both the 1$\sigma$ and 3$\sigma$ upper limits, delineated by solid and dashed lines respectively, on the values of $A_{max}$ versus $T_{\epsilon = 0}$ for both fixing eccentricity to zero and allowing it to vary, for the CLM light curves and deriving stellar parameters from isochrones. We also include the values for previously published detections and upper limits of secondary eclipses in the optical wavelength regime. Also in Figure~\ref{maxalbfig} we plot the cumulative number of systems, and total fraction of all systems, that were modeled in this paper and that have their upper limit of $A_{max}$ below a given value of $A$, for both 1$\sigma$ and 3$\sigma$ upper limits. As can be seen, when fixing eccentricity to zero, we can generally obtain constraints on the maximum possible albedo for $\sim$85\% of the $Kepler$ systems at the 1$\sigma$ confidence level, and $\sim$45\% of systems at the 3$\sigma$ confidence level. When letting eccentricity vary, we can only constrain $A_{max}$ for $\sim$50\% and $\sim$30\% of the $Kepler$ systems at the 1$\sigma$ and 3$\sigma$ confidence levels respectively. However, comparing the $T_{\epsilon = 0}$ values of the previously published planets to that of the $Kepler$ candidates, we find we have significantly increased the number of systems with constrained albedos in the $T_{\epsilon = 0} \lesssim$ 2000 K regime. As can be seen, many of the systems in this temperature regime appear to have maximum possible albedos below 0.3 at the 1$\sigma$ confidence level, thus confirming previous findings of low albedos for hot Jupiters at optical wavelengths, and indicating that such low albedos may be common down to planetary temperatures of 1200~K.

\section{Properties of Some Individual Objects}
\label{indivsec}

In the previous section we analyzed the secondary eclipse detections as a set, in an attempt to find common characteristics of the sample. We have carefully examined each individual system, and in this section present and discuss the most interesting objects in more detail.

\subsection{KOI 1.01 / TrES-2b}

KOI 1.01 is also known as TrES-2b, and was discovered to be a transiting planet by \citet{ODonovan2006} before the $Kepler$ mission was launched. \citet{KippingBakos2011b} determined an upper limit to the eclipse depth of 37ppm at the 1$\sigma$ level, and 73 ppm at the 3$\sigma$ level, based on short-cadence Q0 and Q1 $Kepler$ data, thus limiting the geometric albedo to $A_{g}$ $<$ 0.146 at 3$\sigma$ confidence. \citet{KippingSpiegel2011} have recently published the detection of phase curve variations with an amplitude of 6.5$\pm$1.9 ppm using Q0-Q2 short cadence data, but do not detect the secondary eclipse itself, calculating that any secondary eclipse measurement must have an uncertainty of $\sim$13 ppm. If this variation is due to reflected light, then \citet{KippingSpiegel2011} calculate the the albedo of the planet as Ag = 0.0253$\pm$0.0072.

Using the CLM pipeline light curves, and fixing eccentricity to zero, we determined a secondary eclipse depth of -3.9$^{+8.1}_{-8.0}$ ppm, thus yielding upper limits on the eclipse depth of 4.2 and 20.4 ppm for 1$\sigma$ and 3$\sigma$ confidence levels respectively. We determined a value for the the maximum possible albedo, $A_{max}$, using stellar values from the KIC, of -0.004$^{+0.010}_{-0.019}$, yielding upper limits on the albedo of 0.006 and 0.026 for 1$\sigma$ and 3$\sigma$ confidence levels respectively. If using values from stellar isochrones, we instead determine $A_{max}$ = -0.014$^{+0.028}_{-0.029}$, and thus 1$\sigma$ and 3$\sigma$ upper limits of 0.016 and 0.072, respectively. We did not detect any significant orbital phase variation, although a significant value of the luminosity of the planet must be found in order to produce a significant value of $A_{L_{p}}$ via our modeling technique. Given a difference in the calculated planet-to-star luminosity ratio between our measurements and those of \citet{KippingSpiegel2011} of 10.4$\pm$8.3 ppm, our results do not conflict with those of \citet{KippingSpiegel2011} at a confidence level greater than 1.25$\sigma$.  

We also note that the error on individual points in the Q2 long cadence data is 43 ppm, and thus given the predicted 77.1 minute occultation duration, the 29.4244 minute cadence of $Kepler$ long cadence data, the 88.7 days of coverage, and the 2.47 day orbital period of the system, we calculate that one could detect the secondary eclipse of the planet to a 1$\sigma$ precision of 4.5 ppm. This is in agreement with our 1$\sigma$ upper limit, although our formal 1$\sigma$ errors on the eclipse depth are twice as large, likely due to remaining systematics that were accounted for in the residual-permutation error analysis. However, with better systematic noise reduction, and an additional 1-2 quarters of data, the secondary eclipse of this planet could very well be detected to 3$\sigma$ confidence. Future efforts should be directed towards this goal to confirm the phase signal found by \citet{KippingSpiegel2011}, and ensure it is not due to remaining systematics in the $Kepler$ lightcurves or intrinsic stellar variability.

\subsection{KOI 2.01 / HAT-P-7b}

KOI 2.01 is also known as HAT-P-7 and was discovered by \citet{Pal2008} prior to the launch of the $Kepler$ mission. \citet{Borucki2009} detected a secondary eclipse in the Q0 $Kepler$ data of 130$\pm$11 ppm, and a 122 ppm phase variation. \citet{Christiansen2010} determined an independent 3$\sigma$ upper limit of 550 ppm on the secondary eclipse depth at optical wavelengths using the EPOXI spacecraft. Using Q1 $Kepler$ data, \citet{Welsh2010} determined a secondary eclipse depth of 85.8 ppm, a 63.7 ppm phase variation due to reflection from the planet, a 37.3 ppm phase variation due to ellipsoidal distortions in the star induced by tidal interaction between the planet and star, and determined an albedo in the $Kepler$ passband of 0.18, though no errors on the derived quantities were given.

We do not detect any significant orbital eccentricity ($>$3$\sigma$) in KOI 2.01 / HAT-P-7 in neither the PDC nor the CLM light curves when allowing eccentricity to vary. Fixing eccentricity to zero, for the $Kepler$ PDC light curve, we determine a planet-to-star luminosity ratio of 75.0$^{+9.5}_{-8.1}$ ppm and a value for $A_{L_{p}}$ of 0.421$^{+0.073}_{-0.071}$, and thus a 63.2$^{+13.6}_{-12.6}$ ppm phase variation. Similarly, for the CLM light curve, we derive a planet-to-star luminosity ratio of 77.7$^{+10.3}_{-9.5}$ ppm and a value for $A_{L_{p}}$ of 0.240$^{+0.065}_{-0.045}$, and thus a 37.3$^{+11.2}_{-8.3}$ ppm phase variation. Our determined eclipse depths are very consistent between the CLM and PDC light curves, and within $\sim$1$\sigma$ of the value found by \citet{Welsh2010}, though are not compatible with the value found by \citet{Borucki2009} at the $\sim$4$\sigma$ level. Examining the amplitude of the phase variation of the system, we first point out that there is a $\sim$1.5$\sigma$ difference in the values derived between the PDC and CLM light curves, and likely is a result of the different methods employed to remove systematic noise. While the high-frequency signal of the secondary eclipse was not affected, the low-frequency signal of the phase variation, with a period of $\sim$2.2 days, was much more easily distorted. It is not clear which measurement on the phase variation is more valid, though the CLM pipeline light curves yields a $\chi^{2}_{red}$ of 7.6, versus a value of 26.5 for the PDC data. This result should highlight the level of care that needs to be taken when examining and interpreting phase variations and other low-frequency signals in $Kepler$ data.

Finally, utilizing stellar isochrones for the mass and radius determination of the host star, we determine 3$\sigma$ upper limits to the maximum albedo of the planet of 0.556 and 0.594 for the PDC and CLM light curves respectively, in agreement with the values determined by \citet{Welsh2010}.

\subsection{KOI 10.01 / Kepler-8b}

KOI 10.01 is also known as Kepler-8b, and was first discovered to be a transiting planet by \citet{Jenkins2010c}. \citet{KippingBakos2011a} found that the orbit is consistent with a circular orbit, and placed a 3$\sigma$ upper limit on a secondary eclipse of 101.1 ppm, thus constraining the albedo to $<$ 0.63.

We do not detect any significant eccentricity ($>$3$\sigma$) in neither the PDC nor CLM light curves when allowing eccentricity to vary. Fixing the eccentricity to zero, we place 3$\sigma$ upper limits on the secondary eclipse of the planet at 119 ppm, (14.9$^{+34.7}_{-13.9}$ ppm), and 114 ppm, (19.6$^{+31.6}_{-33.7}$ ppm), for the PDC and CLM light curves respectively. Deriving the parameters for the host star from stellar isochrones yields a 3$\sigma$ limit on the maximum albedo of 0.898 and 0.933 for the PDC and CLM light curves respectively. Thus, we do not provide any additional constraints on the atmosphere of this planet over previous studies.

\subsection{KOI 13.01}

KOI 13 was noted by \citep{Borucki2011} to be a double star, and unresolved in the $Kepler$ images due to the $\sim$4$\arcsec$ pixel size. \citet{Szabo2011} recently conducted a thorough analysis of the system with careful detail to isolating the transiting planet candidate in the double star system, and concluded that KOI 13.01 is likely a brown dwarf with a radius of 2.2$\pm$0.1 $R_{J}$. They also concluded that the transit showed an asymmetrical profile, due to the rapidly rotating nature of the host of KOI 13.01, detected a secondary eclipse with a depth of 120$\pm$10 ppm, and did not find any significant orbital eccentricity. More recently, \citet{Shporer2011} determined a mass of 9.2$\pm$1.1 M$_{Jup}$ via photometric beaming, and \citet{Barnes2011} used the asymmetrical profile of the transit to measure a planetary radius of 1.756$\pm$0.014 R$_{\sun}$, thus making it more likely that this object is a massive hot Jupiter. \citet{Mazeh2011} further support this characterization, and additionally measure a secondary eclipse depth of 163.8$\pm$3.8 ppm.

We do not detect any significant eccentricity ($>$3$\sigma$) in neither the PDC nor CLM light curves when allowing eccentricity to vary, and we can confirm that the asymmetrical transit shape exists in both the PDC and CLM lightcurves. Fixing the eccentricity to zero, we determine a secondary eclipse depth of 124.3$^{+6.9}_{-7.8}$ and 125.6$^{+6.3}_{-8.6}$ ppm for the PDC and CLM light curves respectively, which are statistically consistent with each other and the values found by \citet{Szabo2011}, though is discrepant at the 5$\sigma$ level with the value measured by \citet{Mazeh2011}. We note that KOI 13.01 is the planet with the highest value of $T_{\epsilon=0}$ in our sample, and one of the few that is consistent with a $T_{b}$/$T_{0}$ value of less than 1.0, although we did not take third light into account in our analysis.

\subsection{KOI 17.01 / Kepler-6b}

KOI 17.01 is also known as Kepler-6b, and was discovered to be a transiting planet by \citet{Dunham2010}. \citet{KippingBakos2011a} did not find any evidence for a non-circular orbit, and constrained any possible secondary eclipses to less than 51.5 ppm, and thus a geometric albedo less than 0.32, both at 3$\sigma$ confidence. \citet{Desert2011a} was able to use $Kepler$ data from Q0-Q5, of which Q3-Q5 were not yet publicly accessible at the time of writing, to measure a secondary eclipse of 22$\pm$7 ppm, and did not find any evidence for a non-circular orbit. Combining this eclipse measurement with others obtained via $Spitzer$, they determined a geometric albedo of 0.11$\pm$0.04.

We do not detect any significant eccentricity ($>$3$\sigma$), and place a 3$\sigma$ upper limit on any possible secondary eclipse at 34.5 ppm, (-17.1$^{+17.2}_{-22.3}$ pm). This constrains the planetary albedo, at the 3$\sigma$ level, to less than 0.25 and 0.31 when deriving stellar parameters from the KIC and isochrones respectively, consistent with both the values derived by \citet{KippingBakos2011a} and \citet{Desert2011a}.

\subsection{KOI 18.01 / Kepler-5b}

KOI 18.01 is also known as Kepler-5b, and was discovered to be a transiting planet by \citet{Koch2010}. \citet{KippingBakos2011a} did not find any evidence for a non-circular orbit, though found weak evidence for a secondary eclipse with a depth of 26$\pm$17 ppm, implying a geometric albedo of 0.15$\pm$0.10. \citet{Desert2011a} detected the secondary eclipse, again using Q0-Q5 data, to greater precision and determined a depth of 21$\pm$6 ppm, which they combined with $Spitzer$ observations to determine a geometric albedo of 0.12$\pm$0.04.

We do not detect any significant eccentricity ($>$3$\sigma$), and do not detect the secondary eclipse in neither the PDC nor CLM light curves, placing a 3$\sigma$ upper limit on the eclipse depth of 62.9 ppm (-27.4$^{+30.1}_{-33.2}$ ppm) using the CLM light curve. The derived median value and associated 1$\sigma$ uncertainties on the secondary eclipse depth for the PDC data is 0.21$^{+1.4}_{-0.3}$ ppm, in obvious contradiction to the previously mentioned measured eclipse depths. However, the PDC light curve for KOI 18.01 appears to suffer from a high level of systematic noise, and inspection of the parameter distribution histograms for the surface brightness ratio and luminosity ratio reveal them to significantly deviate from a Gaussian shape, having directly derived 2$\sigma$ uncertainties of $^{+36.1}_{-10.7}$ ppm, thus providing a more reasonable 3$\sigma$ upper limit on the eclipse depth of 54.4 ppm for the PDC light curve.

\subsection{KOI 20.01 / Kepler-12b}

KOI 20.01 has recently been announced by \citet{Fortney2011} as Kepler-12b, a 1.7 $R_{J}$, 0.43 $M_{J}$ planet orbiting a slightly evolved G0 star at a period of 4.4 days. Using $Kepler$ data from Q0-Q7, of which Q3-Q7 were not publicly accessible at the time of writing, they were able to measure a 31$\pm$8 ppm secondary eclipse, which implies a geometric albedo of 0.14$\pm$0.04 when combined with additional $Spitzer$ observations. They also do not detect any significant orbital eccentricity.

Using our CLM light curves, we derived a 3$\sigma$ upper limit on the eclipse depth of 56.3 ppm, (4.7$^{+17.2}_{-9.7}$ ppm), when fixing eccentricity to zero, implying a 3$\sigma$ upper limit on the maximum possible albedo of the planet of 0.40. These results are in agreement with the values recently found by \citet{Fortney2011}.

\subsection{KOI 64.01}

\citet{Borucki2011} noted that KOI 64.01 may be a binary system composed of a F-type primary and a M-type secondary. We do not detect any significant eccentricity ($>$3$\sigma$) in the system, though we do detect marginal evidence for a secondary eclipse in the system. Fixing the eccentricity to zero, we detect a secondary eclipse with a depth of 47.7$^{+25.0}_{-23.3}$ and 75.1$^{+35.1}_{-27.7}$ ppm, (2.0$\sigma$ and 2.7$\sigma$ detections), for the PDC and CLM light curves respectively. Taking the CLM light curve detection as a more reliable measurement, given its $\chi_{red}$ value of 7.3 versus a value of 10.4 for the PDC data, we cannot provide any constraints on the maximum albedo, and in fact even an albedo of 1.0 cannot account for this level of emission. Assuming it has an albedo of 0.0, we calculate a value of $T_{b}$/$T_{0}$ = 1.05$^{+0.29}_{-0.23}$ if deriving stellar values from the KIC, and 1.57$^{+0.09}_{-0.11}$ if using stellar isochrones, which are both above the maximum allowed value of (2/3)$^\frac{1}{4}$ for a planet with no heat redistribution. Thus, this object, if the secondary eclipse detection is real, likely has a significant source of internal energy generation, and certainly may be a brown dwarf or low-mass star. We note however that the effective temperature for the star given in the KIC is 5128 K, which would suggest a K0 spectral type, not F-type. Thus if the host star is a K0 dwarf, the companion, if not a planet, would likely be a brown dwarf, unless the system is composed of a foreground K0 dwarf and a background F+M type eclipsing binary.

\subsection{KOI 97.01 / Kepler-7b}

KOI 97.01 is also known as Kepler-7b, and was discovered to be a transiting planet by \citet{Latham2010}. \citet{KippingBakos2011a} found a secondary eclipse depth of 47$\pm$14 ppm, implying a geometric albedo of 0.38$\pm$0.12, and found a day/night flux difference of 17$\pm$9 ppm. \citet{Demory2011} used $Kepler$ data from Q0-Q4, of which Q3 and Q4 were not yet publicly available at the time of writing, to detect a secondary eclipse of 44$\pm$5 ppm, which implied an albedo of 0.32$\pm$0.03. They also detected an orbital phase curve with an amplitude of 42$\pm$4 ppm. Neither study found evidence for significant orbital eccentricity.

We obtain $>$3$\sigma$ detections of the secondary eclipse in both the PDC and CLM light curves, both fixing eccentricity to zero and allowing it to vary, though we do not detect any significant ($>$3$\sigma$) eccentricity in the system. Fixing the eccentricity to zero, we obtain secondary eclipse depths of 53.2$^{+14.1}_{-13.0}$ ppm and 66.1$^{+17.4}_{-17.5}$ ppm for the PDC and CLM light curves respectively, however we find we cannot place any significant limits on the maximum possible albedo. We determine values for $A_{L_{p}}$ of 0.12$\pm$0.24 and -0.17$^{+0.06}_{-0.10}$ for the PDC and CLM light curves respectively, which translate to phase variations of 12.8$^{+25.8}_{-25.7}$ ppm and -22.5$^{+9.9}_{-14.5}$, neither of which is significant at a $>$3$\sigma$ level. 

Both of our values for the eclipse depth are consistent with those obtained by \citet{KippingBakos2011a} at $<$1$\sigma$ discrepancy, and at $<$1.5$\sigma$ with those of \citet{Demory2011}. Upon inspection of the data, it turns out that we are not able to significantly constrain the albedo, when both \citet{KippingBakos2011a} and \citet{Demory2011} were able to, due to the values we adopt for the stellar mass and radius. We determined median values of 1.09 $M_{\sun}$ and 1.28 $R_{\sun}$ using the KIC, and 1.09 $M_{\sun}$ and 1.03 $R_{\sun}$ via stellar isochrones, where the other studies adopted values of $\sim$1.3 $M_{\sun}$ and $\sim$1.9 $R_{\sun}$, as \citet{Latham2010} found this star to be a G-type sub-giant. Unfortunately the KIC did not hint at this star being non-main-sequence, and the stellar isochrones we employ assume the host star is on the main-sequence. Obviously changing the stellar radius by a factor of $\sim$2 greatly impacts the estimate of reflected light, and this should emphasize the connection between the assumed stellar properties and derived planetary properties. Although we do not detect significant phase variations, our obtained values for the PDC light curve are not in conflict with either previously published result at greater than $\sim$1$\sigma$ significance. The value for the CLM light curve does conflict at $>$3$\sigma$ significance, and we attribute it to differences in the light curve processing from the pixel level data. This is another example of how low-frequency signals can change significantly in $Kepler$ data depending on the reduction technique employed.

\subsection{KOI 183.01}

We highlight KOI 183.01 due to a possibly significant detection of its secondary eclipse and eccentricity. Using the CLM light curves, we obtain a value for the secondary eclipse depth of 14$^{+42}_{-40}$ ppm when fixing eccentricity to zero, but a value of 125$^{+42}_{-39}$, (3.2$\sigma$), when allowing eccentricity to vary. In the latter case, we measure values of $e\cdot$cos($\omega$) = -0.152$^{+0.009}_{-0.008}$ and $e\cdot$sin($\omega$) = 0.03$^{+0.12}_{-0.14}$, yielding values of $e$ = 0.178$^{+0.075}_{-0.023}$ and $\omega$ = 169$^{+47}_{-34}$ degrees.

We employ the Bayesian Information Criterion (BIC) \citep{Schwarz1978} to determine if allowing the eccentricity to vary, thus adding two more free parameters, provides a statistically significantly better fit to the data than fixing it to zero. Given competing models with different values of $\chi^{2}$, and a different number of free parameters, $k$, the value

\begin{equation}
  BIC = \chi^{2} + k\cdot ln(N)
\end{equation}

\noindent is computed, where $N$ is the number of data points, and the model with the lowest BIC value is the preferred model. Given the 3,720 data points in the light curve, 8 free parameters with a $\chi^{2}$ value of 12499 when fixing the eccentricity to zero, and 10 free parameters with a $\chi^{2}$ value of 12476 when allowing eccentricity to vary, values for the BIC of 12565 and 12558 are obtained for the fixed and free eccentricity models respectively. Given the lower BIC value for the eccentricity free model, and given that the resulting parameter distributions from the residual permutation analysis are well-behaved Gaussian curves and do not show any anomalies (see Figure~\ref{ourlcs}.14), we conclude that the eccentricity free model is preferred and statistically significant.

Adopting values for the host star from stellar isochrones, we compute a maximum possible albedo of 0.52$^{+0.21}_{-0.19}$, or a value of $T_{b}$/$T_{0}$ = 1.28$^{+0.08}_{-0.09}$. These are both reasonable values for a moderately reflective planet, one heated beyond radiative thermal equilibrium via tidal heating due to its significant eccentricity, or likely a combination of the two. Additional $Kepler$ data and other follow-up observations should hopefully confirm this detection.

\subsection{KOI 196.01}

KOI 196.01 has recently been confirmed as a 0.49 $M_{J}$, 0.84 $R_{J}$, transiting planet by \citet{Santerne2011} via SOPHIE RV measurements and an analysis of the $Kepler$ light curve. They detect a secondary eclipse with a depth of 64$^{+10}_{-12}$ ppm, with corresponding phase variations, and determine a geometric albedo of 0.30$\pm$0.08. They do not find any significant orbital eccentricity.

We also detect the secondary eclipse with a depth of 77$\pm$24 ppm and 63$^{+33}_{-32}$ ppm in the PDC and CLM light curves respectively, and also do not find any evidence for significant orbital eccentricity, though we do not find any significant orbital phase variation. Via our eclipse depths, and utilizing stellar isochrones, we determine a maximum possible albedo of 0.29$^{+0.10}_{-0.09}$ and 0.26$^{+0.14}_{-0.13}$ for the PDC and CLM light curves respectively. Both of our values for the eclipse depth and geometric albedo are consistent and agree with those determined by \citet{Santerne2011}.

\subsection{KOI 202.01}

We highlight KOI 202.01 due to a possible detection of its secondary eclipse, and a robust upper limit on its albedo. We measure the secondary eclipse depth of the system as 69$^{+31}_{-30}$ ppm (2.3$\sigma$) and 46$^{+31}_{-36}$ ppm (1.3$\sigma$), for the PDC and CLM light curves respectively, and do not detect any significant orbital eccentricity nor phase variations. Although these are not significant enough to claim a robust detection, they are certainly interesting enough results to merit further follow-up with additional data, and place robust 3$\sigma$ upper limits on the maximum albedo of the system at 0.46 and 0.43 for the PDC and CLM light curves respectively, when deriving host star parameters from stellar isochrones.

\subsection{KOI 203.01 / Kepler-17b}

KOI 203.01 is also known as Kepler-17b, and was first confirmed as a transiting planet by \citet{Desert2011b}. Using $Kepler$ Q0-Q6 observations, of which Q3-Q6 were not yet publicly available at the time of writing, along with follow-up observations from $Spitzer$, they were able to detect a secondary eclipse with a depth of 58$\pm$10 ppm, and thus an albedo of 0.10$\pm$0.02, while finding no evidence for any orbital eccentricity. \citet{Bonomo2011} also detect the secondary eclipse at the 2.5$\sigma$ level using Q1-Q2 observations, after fitting and subtracting 4$^{th}$ order polynomials to 4-day segments of the light curve, with a depth of 52$\pm$21 ppm and consistent with a circular orbit.

We were only able to obtain a 3$\sigma$ upper limit on the secondary eclipse depth of 159 ppm (-15$\pm$58 ppm) using the CLM light curve and fixing eccentricity to zero. This corresponds to 3$\sigma$ upper limits on the albedo of 0.26 and 0.28 when deriving the host star parameters from the KIC and stellar isochrones respectively. Our results are thus statistically consistent with the detections of \citet{Desert2011b} and \citet{Bonomo2011}, although our errorbars are much larger. Examining the data, this star's lightcurve has large out of eclipse variations, on the order of the depth of the primary transit, due to both intrinsic stellar variability and systematics introduced by the star's movement, that rendered the PA and PDC data unmodelable. Our CLM pipeline was able to remove a large amount of this variability, (enough to reliably measure the primary transit), but did not fully remove it all, as illustrated by the $\chi_{red}^{2}$ value of 19.4. Thus, the eclipse signal for this planet is below the noise level for the CLM lightcurve.

\subsection{KOI 1541.01}

We highlight this system due to the unusually deep secondary eclipse and high eccentricity we detect at $>$3$\sigma$ confidence. However, the value for the eccentricity, $\sim$0.78, along with the unusually deep eclipse depth of $\sim$1100 ppm, and a large $\chi_{red}^{2}$ value of 28.7, lead us to believe this system suffers from severe systematics which happened to phase together in such a way as to create an artificial eclipse. If the signal is real, then this system must be a background eclipsing binary blend or other similar object.

\subsection{KOI 1543.01}

This system is very similar to KOI 1541.01 in that we also obtain a $>$3$\sigma$ detection of a secondary eclipse, but with an unusually high values for the eclipse depth, eccentricity, and $\chi_{red}^{2}$. Inspection of the light curve also reveals this system to contain significant systematics, or else must be a background eclipsing binary.

\section{Summary and Conclusions}
\label{concsec}

We have analyzed the $Kepler$ Q2 light curves of 76 hot Jupiter transiting planet candidates using both the $Kepler$ PDC data and the results from our own photometric pipeline for producing light curves from the pixel-level data. Of the 76 initial candidates only 55 have light curves with high enough photometric stability to search for secondary eclipses. For the remaining targets this search is hindered by either intrinsic variability of the host star or residual systematics in the light curve analyses. We have found that significant systematics in the $Kepler$ light curves due to small photometric apertures and large centroid motions hinder analyses if not properly removed or accounted for, and that a re-reduction of the photometry is best done at the pixel-stamp level. We also stress the importance of taking into account how $Kepler$ light curves are produced from the pixel level data when considering any detection of low-frequency signals, as they can vary significantly depending on the technique employed. 

We have also re-determined the stellar and planetary parameters of each system while deriving robust errors that take into account residual systematic noise in the light curves. We detect what appear to be the secondary eclipse signals of $\sim$20-30 of the targets in our list at $> 1 \sigma$ confidence levels, and also derive robust upper limits for the secondary eclipse emission of all the remaining objects in our sample. All of our sample present excess emission compared to what is expected via blackbody thermal emission alone, as well as a trend of increasing excess emission with decreasing expected maximum effective planetary temperature, in agreement with previously reported secondary eclipse detections of hot Jupiters in the optical, which can be attributed to the 'appearance' of increasing albedos with decreasing planetary temperatures. By performing statistical analyses of those results we arrive to the following main conclusions.

\begin{enumerate}

\item Assuming no contribution from reflected light, i.e., A = 0.0, the majority of the detected secondary eclipses reveal thermal emission levels higher than the maximum emission levels expected for planets in local thermodynamical equilibrium.

\item While the extra emission from many of the planets can be accounted for by varying the amount of reflected light, the emission from  $\sim$50\% of the detected objects ($>$1$\sigma$) can not be accounted for even when assuming perfectly reflective planets (i.e., A = 1.0). These planets must either have much higher thermal emission in the optical compared to a blackbody, as predicted by theoretical models, have very large non-LTE optical emission features, have underestimated host star masses, radii, or effective temperatures, or are in fact false-positives and not planets but rather brown dwarfs, very low-mass stars, or stellar blends. Follow-up observations of these systems are necessary to confirm this conclusion. The most outstanding potentially false positive systems are KOI 64.01, 144.01, 684.01, 843.01, 1541.01, and 1543.01. Given that this is 6 of the 55 systems that we modeled, or 11\% of the sample, it appears to agree with the expected $\sim$10\% false positive rate of the initial 1,235 candidates estimated by \citet{Morton2011}.

\item Although we do not identify a sole cause of the observed trend of increasing excess planetary emission with decreasing expected maximum effective planetary temperature, the hypothesis of increasing planetary albedo with decreasing planetary temperature is able to explain many of the systems. This would be physically plausible as the upper atmospheres of Jupiter-like planets transition from very low albedos at high temperatures, as observed for hot Jupiters, to higher albedos at lower temperatures, as observed for cool Jupiters in our own solar system. We note that further observations via Kepler and multi-wavelength ground-based facilities of both the planetary candidates and their host stars are still needed to fully explain this trend.

\item From the emission upper limits placed on planet candidates for which we do not detect secondary eclipse signals, we conclude that a significant number, at least 30\% at the 1$\sigma$ level, of those targets must have very low-albedos, ($A_{g}$ $<$ 0.3), which is a result that is consistent with the majority of previous observations and early theoretical hot Jupiter model predictions. All previous observational upper limits had been placed on hot Jupiters with expected atmospheric effective temperatures higher than $\sim$1650 K. Our results extend that temperature limit to planets with expected effective temperatures higher than 1200 K and can help further establish what chemicals place a critical role in the atmospheric properties of hot Jupiters.

\item From the inspection of individual targets we conclude that the majority of our secondary eclipse depths for candidates with previously published eclipse detections are consistent with the results from those other studies. We note that several of those other studies have access to $Kepler$ data from quarters after Q2, while our analysis is limited to just the public Q2 data, and therefore our results have larger detection errorbars in some cases.

\item Our results are based on only one quarter of $Kepler$ data, but 12-24 quarters should become available in the future. Therefore, we expect that future studies of these targets will be able to improve our secondary eclipse detections and upper limits by factors of 3-5, or much greater if noise systematics in the light curves can be further reduced.

\end{enumerate}

\acknowledgments
We first and foremost thank the entire \emph{Kepler} team and all those who have contributed to the \emph{Kepler Mission}, without which this paper would not be possible. We thank the referee for a very thorough and truly helpful report, which greatly improved the manuscript. J.L.C acknowledges support through a National Science Foundation Graduate Research Fellowship. J.L.C. \& M.L.M are grateful for funding from \emph{Kepler's} Guest Observer Program. Some/all of the data presented in this paper were obtained from the Multimission Archive at the Space Telescope Science Institute (MAST). STScI is operated by the Association of Universities for Research in Astronomy, Inc., under NASA contract NAS5-26555. Support for MAST for non-HST data is provided by the NASA Office of Space Science via grant NNX09AF08G and by other grants and contracts. This research has made use of NASA's Astrophysics Data System.

\bibliography{ref.bib}

\clearpage
\LongTables
\begin{deluxetable*}{rcccccccccc}
  \tablecaption{Modeling Results: Median Values and Associated 1$\sigma$ Uncertainties}
  \tablecolumns{11}
  \tablehead{KOI & $J$ & $r_{\rm sum}$ & $k$ & $i$ & $e$cos$w$ & $e$sin$w$ & $P$ & $T_{0}$ & $A_{L_{p}}$ & $\chi^{2}_{red}$\\ & & & & ($\degr$) & & & (Days) & (BJD-2450000) & & }
  \startdata
  \cutinhead{PDC Light Curve With Eccentricity Fixed to Zero}
     2.01 & $ $ 0.0108$^{_{+0.0013}}_{^{-0.0011}}$ & 0.254$^{_{+0.005}}_{^{-0.005}}$ & 0.0769$^{_{+0.0004}}_{^{-0.0004}}$ & 83.92$^{_{+0.7045}}_{^{-0.5848}}$ & $ $ 0.000$^{_{+0.000}}_{^{-0.000}}$ & $ $ 0.000$^{_{+0.000}}_{^{-0.000}}$ & 2.204732$^{_{+3.1\textrm{e-}06}}_{^{-3.0\textrm{e-}06}}$ & 4954.35796$^{_{+0.00013}}_{^{-0.00013}}$ & $ $ 0.421$^{_{+0.07}}_{^{-0.07}}$ & 26.5\\
   5.01 & $ $-0.0088$^{_{+0.0088}}_{^{-0.0097}}$ & 0.123$^{_{+0.010}}_{^{-0.013}}$ & 0.0356$^{_{+0.0009}}_{^{-0.0011}}$ & 83.60$^{_{+0.7978}}_{^{-0.5911}}$ & $ $ 0.000$^{_{+0.000}}_{^{-0.000}}$ & $ $ 0.000$^{_{+0.000}}_{^{-0.000}}$ & 4.780381$^{_{+6.7\textrm{e-}05}}_{^{-6.7\textrm{e-}05}}$ & 4965.97227$^{_{+0.00119}}_{^{-0.00121}}$ & $ $ 0.300$^{_{+0.72}}_{^{-0.35}}$ & 2.69\\
  10.01 & $ $ 0.0014$^{_{+0.0033}}_{^{-0.0013}}$ & 0.151$^{_{+0.004}}_{^{-0.004}}$ & 0.0938$^{_{+0.0006}}_{^{-0.0007}}$ & 84.68$^{_{+0.3495}}_{^{-0.2975}}$ & $ $ 0.000$^{_{+0.000}}_{^{-0.000}}$ & $ $ 0.000$^{_{+0.000}}_{^{-0.000}}$ & 3.522511$^{_{+1.5\textrm{e-}05}}_{^{-1.6\textrm{e-}05}}$ & 4954.11838$^{_{+0.00043}}_{^{-0.00041}}$ & $ $-5.752$^{_{+4.04}}_{^{-64.5}}$ & 4.31\\
  13.01 & $ $ 0.0248$^{_{+0.0013}}_{^{-0.0015}}$ & 0.294$^{_{+0.005}}_{^{-0.006}}$ & 0.0659$^{_{+0.0002}}_{^{-0.0003}}$ & 79.79$^{_{+0.5070}}_{^{-0.4079}}$ & $ $ 0.000$^{_{+0.000}}_{^{-0.000}}$ & $ $ 0.000$^{_{+0.000}}_{^{-0.000}}$ & 1.763589$^{_{+2.3\textrm{e-}06}}_{^{-2.3\textrm{e-}06}}$ & 4953.56510$^{_{+0.00013}}_{^{-0.00014}}$ & $ $ 0.339$^{_{+0.02}}_{^{-0.02}}$ & 15.9\\
  18.01 & $ $ 0.0000$^{_{+0.0001}}_{^{-0.0000}}$ & 0.174$^{_{+0.027}}_{^{-0.005}}$ & 0.0783$^{_{+0.0025}}_{^{-0.0007}}$ & 87.64$^{_{+1.7674}}_{^{-3.6176}}$ & $ $ 0.000$^{_{+0.000}}_{^{-0.000}}$ & $ $ 0.000$^{_{+0.000}}_{^{-0.000}}$ & 3.548460$^{_{+2.4\textrm{e-}05}}_{^{-2.5\textrm{e-}05}}$ & 4955.90081$^{_{+0.00068}}_{^{-0.00064}}$ & $ $-78.785$^{_{+238.}}_{^{-525.}}$ & 9.33\\
\enddata
  \label{modelresultstab}
\end{deluxetable*}

\clearpage

\begin{deluxetable*}{rrcccccccc}
  \tablewidth{0pt}
  \tabletypesize{\tiny}
  \tablecaption{Derived System Parameters and Associated 1$\sigma$ Uncertainties}
  \tablecolumns{10}
  \tablehead{KOI & $L_{r}$ & $\sigma_{sec}$ & $T_{\rm 0}$ & $T_{\epsilon=0}$ & $T_{\rm b}$\tablenotemark{1} & $T_{\rm b}$/$T_{0}$ & $R_{\rm p}$ & $a$ & $A_{\rm max}$\\ & (ppm) & & (K) & (K) & (K) & & ($R_{\rm Jup}$) & (AU) & } 
  \startdata
  \cutinhead{PDC Light Curve with Eccentricity Fixed to Zero and Stellar Parameters from the KIC}
     2.01 & $ $   75.0$^{_{+9.48}}_{^{-8.13}}$ & $ $9.11 & $ $2702$^{_{+0759}}_{^{-0597}}$ & $ $2441$^{_{+0686}}_{^{-0540}}$ & $ $ 2851$^{_{+0062}}_{^{-0064}}$ & $ $ 1.055$^{_{+0.294}}_{^{-0.232}}$ & 1.022$^{_{+0.955}}_{^{-0.497}}$ & 0.0367$^{_{+0.0132}}_{^{-0.0097}}$ & $ $ 0.38$^{_{+0.65}}_{^{-0.24}}$  \\
   5.01 & $ $  -13.4$^{_{+13.4}}_{^{-14.8}}$ & $ $-0.99 & $ $2139$^{_{+0597}}_{^{-0455}}$ & $ $1932$^{_{+0539}}_{^{-0411}}$ & $ $-2607$^{_{+0827}}_{^{-0256}}$ & $ $-1.159$^{_{+0.539}}_{^{-0.389}}$ & 0.617$^{_{+0.552}}_{^{-0.300}}$ & 0.0586$^{_{+0.0205}}_{^{-0.0154}}$ & $ $-0.40$^{_{+0.39}}_{^{-1.15}}$  \\
  10.01 & $ $   14.9$^{_{+34.7}}_{^{-13.9}}$ & $ $1.07 & $ $1993$^{_{+0546}}_{^{-0420}}$ & $ $1801$^{_{+0493}}_{^{-0380}}$ & $ $ 2319$^{_{+0205}}_{^{-0358}}$ & $ $ 1.121$^{_{+0.355}}_{^{-0.321}}$ & 0.990$^{_{+0.905}}_{^{-0.470}}$ & 0.0471$^{_{+0.0172}}_{^{-0.0124}}$ & $ $ 0.21$^{_{+0.55}}_{^{-0.17}}$  \\
  13.01 & $ $  124.3$^{_{+6.93}}_{^{-7.81}}$ & $ $16.15 & $ $5069$^{_{+1414}}_{^{-1135}}$ & $ $4580$^{_{+1278}}_{^{-1025}}$ & $ $ 3744$^{_{+0055}}_{^{-0056}}$ & $ $ 0.738$^{_{+0.213}}_{^{-0.161}}$ & 1.610$^{_{+1.524}}_{^{-0.779}}$ & 0.0348$^{_{+0.0127}}_{^{-0.0092}}$ & $ $ 0.23$^{_{+0.40}}_{^{-0.14}}$  \\
  18.01 & $ $    0.2$^{_{+1.42}}_{^{-0.27}}$ & $ $0.33 & $ $1827$^{_{+0514}}_{^{-0398}}$ & $ $1651$^{_{+0464}}_{^{-0360}}$ & $ $ 1648$^{_{+0132}}_{^{-0252}}$ & $ $ 0.866$^{_{+0.278}}_{^{-0.267}}$ & 0.749$^{_{+0.703}}_{^{-0.357}}$ & 0.0446$^{_{+0.0158}}_{^{-0.0118}}$ & $ $ 0.01$^{_{+0.03}}_{^{-0.01}}$  \\
  64.01 & $ $   47.7$^{_{+25.0}}_{^{-23.3}}$ & $ $2.05 & $ $2711$^{_{+0748}}_{^{-0605}}$ & $ $2449$^{_{+0676}}_{^{-0547}}$ & $ $ 2702$^{_{+0165}}_{^{-0227}}$ & $ $ 0.980$^{_{+0.294}}_{^{-0.226}}$ & 0.829$^{_{+0.756}}_{^{-0.403}}$ & 0.0324$^{_{+0.0115}}_{^{-0.0084}}$ & $ $ 0.24$^{_{+0.49}}_{^{-0.17}}$  \\
\enddata
  \tablenotetext{1}{Derived assuming $A$ = 0.0}
  \label{tab3}
\end{deluxetable*}

\end{document}